\documentclass[12pt]{article}
\usepackage{epsfig}
\usepackage{graphicx}
\usepackage{cite}
\usepackage{amsfonts}
\usepackage{amssymb}
\usepackage{bm}
\usepackage{slashed}
\usepackage{latexsym}
\usepackage{amsmath}
\numberwithin{equation}{section}
\def\ee{\end{equation}}
\def\ba{\begin{eqnarray}}
\def\ea{\end{eqnarray}}

\def\bq{\begin{quote}}
\def\eq{\end{quote}}

 at 10truept
\newcommand{\beq}{\begin{equation}}
\newcommand{\eeq}{\end{equation}}
\newcommand{\beqa}{\begin{eqnarray}}
\newcommand{\eeqa}{\end{eqnarray}}
\newcommand{\bea}{\begin{eqnarray}}
\newcommand{\eea}{\end{eqnarray}}

\newcommand{\ex}[1]{\noindent{\bf Exercise:} {\it #1}}

\newcommand{\hf}{\frac{1}{2}}

\def\ltap{\ \raise.3ex\hbox{$<$\kern-.75em\lower1ex\hbox{$\sim$}}\ }
\def\gtap{\ \raise.3ex\hbox{$>$\kern-.75em\lower1ex\hbox{$\sim$}}\ }
\def\gl{\ \raise.5ex\hbox{$>$}\kern-.8em\lower.5ex\hbox{$<$}\ }
\def\roughly#1{\raise.3ex\hbox{$#1$\kern-.75em\lower1ex\hbox{$\sim$}}}
\def\calo{{\cal O}}
\def\calh{{\cal H}}
\def\bv{{\bf b}}
\def\semi{;\hfil\break}

\begin{document}

\thispagestyle{empty}
\begin{titlepage}
\nopagebreak

\title{
The gravitational S-matrix: Erice lectures\footnote{Presented at the 48th Course of the Erice International School of Subnuclear Physics, ``What is known and unexpected at LHC," Aug./Sept. 2010.}} 
\vfill
\author{Steven B. Giddings\footnote{giddings@physics.ucsb.edu}}
\date{ }

\maketitle

{\it  Department of Physics, 
University of California, Santa Barbara, CA 93106}
\vskip 0.5cm

\vfill
\begin{abstract}
These lectures discuss an S-matrix approach to quantum gravity, and its relation to more local spacetime approaches.  Prominent among the problems of quantum gravity are those of unitarity and observables.  In a unitary theory with solutions approximating Minkowski space, the S-matrix (or, in four dimensions, related inclusive probabilities) should be sharply formulated and physical.  Features of its perturbative description are reviewed.  A successful quantum gravity theory should  in particular address the questions posed by the ultrahigh-energy regime.  Some control can be gained in this regime by varying the impact parameter as well as the collision energy.  However, with decreasing impact parameter gravity becomes strong, first eikonalizing, and then entering the regime where  in the classical approximation black holes form.  Here one confronts what may be the most profound problem of quantum gravity, that of providing unitary amplitudes, as seen through  the information problem of black hole evaporation.  Existing approaches to quantum gravity leave a number of unanswered questions in this regime; there are strong indications that new principles and mechanisms are needed, and in particular there is a good case that usual notions of locality are inaccurate.  One approach to these questions is investigation of the approximate local dynamics of spacetime, its observables, and its limitations; another is to directly explore properties of the gravitational S-matrix, such as analyticity, crossing, and others implied by gravitational physics.  

 \end{abstract}
 \vskip.4in
 
\noindent
\vfill

\vfill
\end{titlepage}

\section{The problem(s) of quantum gravity}

Reconciliation of quantum mechanics with gravity may well be the most profound theoretical problem left unsolved by twentieth-century physics.  Both general relativity and quantum-mechanics became well established in the first quarter of the last century, yet their clash continues to this day.  One might simply declare the problem too difficult, and move on to more modest goals.  But, there are glimmers of hope, both from new approaches to the problem, and perhaps more importantly, through a better understanding of the  nature of the underlying difficulty.  

There are a number of sub-problems to that of quantizing gravity.  A list includes: 

\begin{enumerate}

\item The problem of UV divergences and nonrenormalizability:  what structure cures an apparent infinite proliferation of perturbative ultraviolet divergences, and needed counterterms?

\item The problem of singularities:  how can a quantum theory resolve singularities of the classical theory, such as are found in black holes and in many cosmologies?

\item The problem of observables/time:  how can one formulate gauge invariant observables in the quantum theory, given that diffeomorphisms are gauge transformations in the classical theory?

\item The problem of unitarity: what dynamics unitarizes amplitudes, in the high-energy regime?

\item The conundrums of cosmology: this includes the problem of defining infrared finite observables in inflationary cosmology -- an extreme example is the measure problem; or, perhaps we need an alternative to inflation.

\end{enumerate}

Historically, there has been a lot of focus on problems one and two, and they have strongly motivated approaches to quantum gravity such as string theory and loop quantization.  However, there are increasing indications that problems three through five are both more profound, and are linked; they seemingly drive more deeply at the heart of the central problem of quantum gravity.  

Indeed, there has been a growing appreciation of the fact that there are deep {\it long-distance} issues in quantum gravity, which do not appear easily resolved by short-distance modifications of the theory.  This suggests that a successful theory must incorporate more radical departures from local quantum field theory (QFT), relevant over even very long-distance scales, in certain circumstances.  These lectures will explore one way to probe these questions -- through study of the gravitational S-matrix.

\section{The gravitational S-matrix}

\subsection{Observables in quantum gravity?}
\label{obssubsec}

Given our difficulties with gravity, it is important to understand what are sharply formulated questions.  One might na\"\i vely expect that a theory of quantum gravity should calculate correlation functions -- as in QFT -- such as
\beq
\langle \phi(x) \phi(y)\cdots\rangle\ ,\ \langle g_{\mu\nu}(x)\rangle\ , \ \langle \phi(x) \phi(y) g_{\mu \nu}(z)\cdots\rangle\ ,\quad {\rm etc.}\ ,
\eeq
where $\phi$ is some matter field, and $g_{\mu\nu}$ is the metric.  However, such correlators cannot be gauge invariant, and hence physical.  The reason is that the gauge symmetries of gravity, at least at the semiclassical level, are diffeomorphisms, whose infinitesimal forms act on local fields by
\beq
\delta \phi(x) = \xi^\mu \partial_\mu\phi(x)\ ,\ \delta g_{\mu\nu}(x) = \nabla_{(\mu}\xi_{\nu)}(x)\ , {\rm etc.}
\eeq
In field theory, local observables are given in terms of the fundamental fields; but in gravity, by this reasoning, no local observables can be gauge invariant.

Clearly our actual observations are governed by a theory incorporating both quantum mechanics and gravity, and we apparently make at least approximately local observations, certainly as compared to cosmological scales -- so how is this problem resolved?  A proposed answer goes back to Leibniz and Einstein, and began to be concretely developed by DeWitt: an important class of observables in gravity are {\it relational}, that is give location with relation to other features of the particular quantum state of the system.  In particular, \cite{GMH} explored formulation, within a semiclassical/perturbative approximation, of a class of {\it protolocal} observables, which are relational and {\it approximately} reduce to local observables in certain states.  Very concrete examples of this construction can be given, in the context of two-dimensional gravity, using world-sheet methods of string theory\cite{GaGiobs}. Other approaches to implementing relational ideas have been explored in a number of places in the literature;\footnote{For one list of references, see \cite{GMH}.}  moreover, such relational ideas are concretely used to handle gauge invariance in inflationary cosmology, where, for example, fluctations are computed on time slices defined by the inflaton field taking on a particular value, like that corresponding to ``reheating."

While relational observables are important for quantum cosmology, there are many subtleties in their proper definition, which ultimately appears to require a deeper knowledge of the quantum theory, and also encounters various infrared issues.  For that reason, in the present discussion I will primarily  focus on another approach to formulation of sharp physical quantities:  the {\it gravitational S-matrix}.\footnote{Important early references on gravitational scattering include \cite{tHoo,MuSo,ACV,ACVCQG,Verlinde:1991iu,BaFi}.}

\subsection{The S-matrix and the ultraplanckian regime}

A starting point for discussing the gravitational S-matrix is the observation that, to a very good approximation, Minkowski space looks like a good solution of the theory of quantum gravity.  Indeed, the curvature radius of the observed Universe is of order $10^{60}$ in Planck units -- flat space is an excellent approximation for all but the longest-scale questions, and we therefore assume that it makes sense to consider the exactly flat limit.  

Next, we also observe that there are excitations about this which we describe as ``particle" states, such as electrons, photons, {\it etc.}, whatever their more fundamental description may be.  Their asymptotic states are described by their momenta $p_i$, and other quantum numbers.  Moreover, we can consider asymptotic multi-particle states, consisting of widely separated particles.  And, in a quantum theory, we can ask for the amplitudes to transition from a given such ``in" state, in the far past, to another ``out" state, in the far future.  This collection of amplitudes,  which we might denote
\beq
S(p_{i'}, p_i) = {}_{out}\langle p_{i'}| p_i\rangle_{in}\ ,
\eeq
is the S-matrix.  

I should note there are two technicalities that we are glossing over.  First, for gravity in $D=4$ spacetime dimensions, these amplitudes aren't well-defined: instead we must perform a sum over soft gravitons, like with soft photons in QED.  However, as we will see, this technicality can be avoided in higher dimensions.  Secondly, a more precise definition of the S-matrix involves a map between normalizable Hilbert-space states; momentum-space states give a singlular limit of these, though one whose handling is well-understood in asymptotically flat spacetimes.

Such an S-matrix, which is tightly constrained by properties such as unitarity and analyticity, can be a very powerful way to summarize our ignorance of a theory.  The S-matrix approach is so powerful that it lead to the {\it discovery} of string theory: Veneziano\cite{Venstdisc} guessed amplitudes satisfying  additional ``duality" properties, and they were then found  to describe excitations of one-dimensional objects.  We might anticipate that such study in the context of gravity, supplemented by additional physical input, could bear important fruit.  

There are multiple proposed approaches to quantizing gravity, and the S-matrix also provides a very important test for these:  any  theory of quantum gravity should give us a means to calculate $S$, at least approximately, or must provide us with some alternative physical quantities.  

This test becomes particularly incisive in the ultraplanckian regime, at center of mass energies $E\gg M_D$, where $M_D$ denotes the D-dimensional Planck mass.  That any theory of quantum gravity must describe this regime follows from very general principles, and  avoiding this regime appears to require radical assumptions.  The first principle is that of Lorentz invariance: the Minkowski-space solution is invariant under boosts, and we can perform an arbitrarily large boost on a single-particle excitation of it to get a state with an arbitrarily large energy.\footnote{There are advocates of Lorentz-invariance violation.  This won't be considered here because a) Lorentz invariance is hard to violate by a small amount; b) there are very stringent constraints on such violation; c) it is hard to reconcile such violation with basic properties of black holes, {\it e.g.} in the astrophysical context; and d) if black holes are still possible, they would form in other contexts, {\it e.g.} that of high-mass collapse, and we would still face similar puzzles to those I will describe.}  The second principle is a very weak notion of locality:  one can consider independently boosted particles at very large separations, and so prepare a state with a large center of mass energy -- and the then allow the particles to collide.

Plausibly during the big bang particles reached such energies, and it is interesting to ask whether such collisions are likely to take place anywhere in nature today.  We observe that cosmic ray accelerators (argued to be active galactic nuclei) produce particles up to $\sim10^{12}$ GeV, and now and then they could collide; that is only short by ${\cal O}(10^7)$.  Even more interestingly, in scenarios with large and/or strongly warped extra dimensions, the fundamental Planck mass $M_D$ could be as low as the TeV range, and in the most optimistic scenarios black holes could give important signatures at LHC\cite{GiTh,DiLa}!  While I won't talk about this phenomenology, there are multiple reviews, including \cite{Pascos}.

Whether or not this physics is experimentally accessible, the gedanken experiment and theoretical problem remain: we need a calculational framework that makes predictions in the ultraplanckian regime.  As I will describe, here an apparently critical and fundamental problem is {\it unitarity}.  However, as viewed through the lens of earlier attempts to quantize gravity, a more basic concern is that of {\it nonrenormalizability}.  We will begin to understand this problem, and then explore the apparently more fundamental issue of unitarity, starting with the perturbative approach to gravity.

\section{Review of perturbative gravity}
\label{Pertquant}
\subsection{Perturbative quantization and the Born amplitude}

A starting point for perturbative quantization is the action,
\beq\label{gravact}
S= \int d^D x \sqrt{-g}\left[{1\over 16\pi G_D} {\cal R} +{\cal L}_M\right]
\eeq
where $\cal R$ is the Ricci scalar, ${\cal L}_M$ the matter lagrangian, and Newton's constant $G_D$ is related to $M_D$ by the {\it Particle Data Book}\cite{PDB} convention:
\beq
M_D^{D-2}={(2\pi)^{D-4} / (8 \pi G_D) }\ .
\eeq
For simplicity we will typically consider for matter  a minimally-coupled scalar of mass $m$,
\beq
{\cal L}_m = -\frac{1}{2}\left(\nabla\phi\right)^2-\frac{1}{2}m^2\phi^2\ .
\eeq
We investigate perturbations about Minkowski space, 
\beq
g_{\mu\nu}= \eta_{\mu\nu} + \sqrt{32\pi G_D}\, h_{\mu\nu}\ 
\eeq
and expand in powers of $h$.  This gives an action of the form
\beq\label{pertact}
S= S_{m,0} + \int d^D x\left\{ {1\over 2} h^{\mu\nu}({\hat {\cal L}}h)_{\mu\nu} +\sqrt{8\pi G_D}[ h^{\mu\nu}T^{m,0}_{\mu\nu}+ {\cal L}_3(h)]+\cdots\right\}\ .
\eeq
Here, quantities evaluated with $g=\eta$ are labeled with zeros, $\hat {\cal L}$ is a second-order differential operator (a tensor analog of the laplacian), indices are raised and lowered with $\eta$, $T$ is the stress tensor, and ${\cal L}_3$ is the third-order term in $h$.  The ellipses contain higher powers of $\phi$ and $h$.  The propagators are given by inverting the kinetic operators for $\phi$ and $h$; the latter requires gauge fixing.  There are two approaches to this, both using the quantity
\beq
{\bar h}_{\mu\nu} = h_{\mu\nu} -\hf \eta_{\mu\nu} h^\lambda_\lambda\ .
\eeq
The first is to apply the gauge fixing condition 
\beq
\partial^\mu {\bar h}_{\mu\nu}=0\ .
\eeq
Alternately, one may add a gauge fixing term
\beq
{\cal L}_{gf} = {1\over 16\pi G_D \alpha} \left(\partial^\mu {\bar h}_{\mu\nu}\right)^2
\eeq
to the lagrangian; particularly nice simplifications occur for $\alpha=-1$.  In either case, one straightforwardly finds the propagator, and using that can immediately calculate a first amplitude, that for tree-level scattering of two matter particles.  This is the Born amplitude.  It is most conveniently given in terms of the Mandelstam invariants; let $p_1$, $p_2$ denote the momenta of the incoming particles, and $-p_3$, $-p_4$ those for outgoing, and define
\beq
s=-(p_1+p_2)^2=E^2\ ,\ t=-(p_3+p_1)^2 = -q^2\ ,\ u = -(p_4+p_1)^2\ .
\eeq
Here $E$ is the total center-of-mass (CM) energy, $q$ is the ``momentum transfer," and one finds $s+t+u=4m^2$.
If we then define the reduced transition matrix element by 
\beq
S=1+iT(s,t)(2\pi)^D\delta^D(\sum p)\ 
\eeq
and take the limit of large $s=E^2$, we find the high-energy approximation of the Born amplitude $T_0$:
\beq\label{bornamp}
T_0(s,t) \simeq -8\pi G_D  s^2/t\ .
\eeq

\ex{Derive this formula from the Feynman rules of the action (\ref{pertact}).}

\begin{figure}[!hbtp]
\begin{center}
\includegraphics[width=6cm]{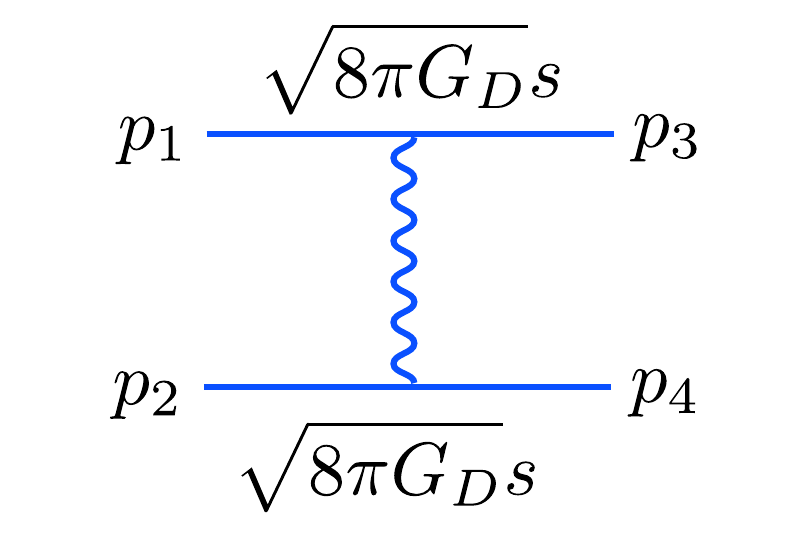}
\caption{A Feynman diagram representing scattering in the Born approximation.}\label{fig:born}
\end{center}
\end{figure} 

This formula is easy to understand; see fig.~\ref{fig:born}.  There is a coupling constant $\sim \sqrt{8\pi G_D}s$ at each vertex, and the graviton propagator gives a pole $\sim 1/q^2$.  
So, we're on our way to computing the S-matrix!  The question is where $T_0$ gives a good approximation to the full S-matrix.  As I will explain further, the answer is apparentely for small enough $t$, or equivalently, large enough impact parameter $b$.  To see this, we need to consider possible corrections to (\ref{bornamp}), from loops or radiation.

\subsection{Radiative corrections, nonrenormalizability, and the problem of ultraplanckian energies}

\begin{figure}[!hbtp]
\begin{center}
\includegraphics[width=6cm]{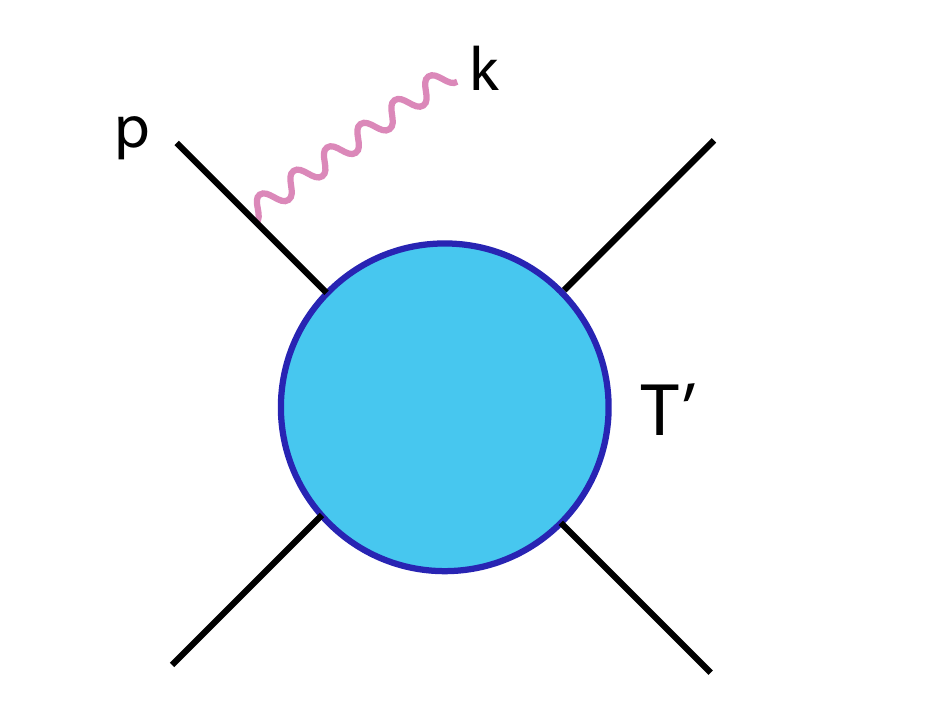}
\caption{Addition of a soft graviton to a diagram.}\label{fig:soft}
\end{center}
\end{figure}

First, note that an accelerated particle will radiate gravitational bremsstrahlung,  analogous to that of QED.  To estimate how it corrects a given process, say with amplitude $T'$, we attach a graviton of momentum $k$ to one of the legs of $T'$, as in fig.~\ref{fig:soft}.  Then there is an additional external propagator factor, and the amplitude becomes
\beq
T_{soft} \propto {T'\over (p-k)^2+m^2 } = -{T'\over 2p\cdot k}
\eeq
where we use the on-shell condition for the leg's momentum $p$, and for the graviton momentum.  The sum over final graviton states thus gives
\beq
\int d^D k\,\delta(k^2)\, |T_{soft}|^2 \sim \int {d^D k\over k^4}\ .
\eeq

\ex{Work out the details of this calculation, checking the form of the preceding expression.}

This is infrared divergent for $D\leq 4$, and is an indicator of the soft-graviton problem alluded to before.  Like with QED, the S-matrix will not be defined, but the resolution appears straightforward\cite{WeinSG}:  include IR divergences from loop corrections, and perform an inclusive sum over scattering probabilities with gravitons below a momentum corresponding to, {\it e.g.}, experimental resolution, to get an IR finite result.  For $D>4$ there is no such IR divergence obstructing the definition of S, and rough estimates\cite{GiSr} indicate that the emission probabilities are small at small $t$/large $b$.

Next consider loops.  Since the coupling constant $G_D$ has negative mass dimension, there will be UV divergences requiring counterterms of the form
\beq
{\Delta{\cal L}}\propto G_D^a p^b h^c M^d\ ,
\eeq
where $M$ is a UV cutoff scale, and the exponents satisfy $a(D-2)+D=b+{c\over 2}(D-2)+d$.  Thus, as one considers arbitrarily high-order loops (increasing $a$), there can be arbitrarily many different higher-dimension counterterms, unless there is some new magic to prevent them.  The theory is {\it nonrenormalizable}, and really this means nonpredictive.  Once scattering energies approach the Planck scale, amplitudes can depend on all of these counterterms, and thus there is no finite set of measurements which can determine the theory.

This of course does not mean that the action (\ref{gravact}) is useless!  Instead, we  take the viewpoint of effective field theory:  the Einstein action makes good (and in some cases, very well-tested) predictions in ``low energy" regimes, for example in scattering at $E\ll M_D$.  The loop corrections indeed renormalize $G_D$ and the cosmological constant $\Lambda$, which must be adjusted to match their observed low-energy values.  But, corrections corresponding to higher-dimension operators, such as ${\cal R}^2$, are {\it irrelevant}, that is, make negligible contribution at low energy.  In particular, the Born amplitude remains a good approximation for scattering in this regime.

\begin{figure}[!hbtp]
\begin{center}
\includegraphics[width=15cm]{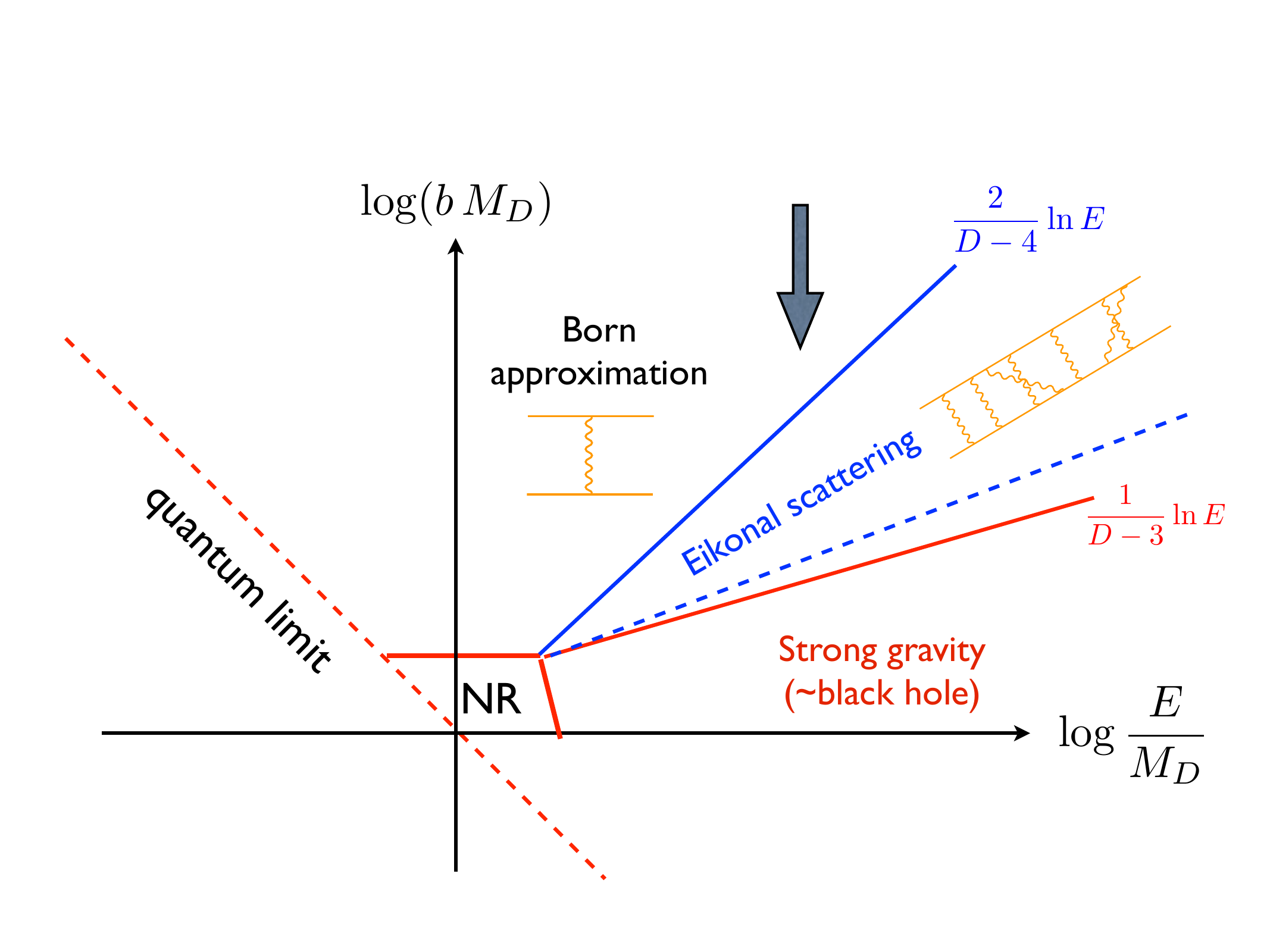}
\caption{A proposed ``phase diagram" of different regimes for gravitational scattering.  In particular, we consider the effect of decreasing impact parameter, at fixed ultraplanckian energy, as indicated by the arrow.  NR indicates the regime where higher-dimension operators are expected to be important.}\label{Ebdiag}
\end{center}
\end{figure}

As $E$ approaches $M_D$, we at first suspect that nonrenormalizability is severely problematical, and that we lose ability to say anything about scattering.  However, a little reflection raises a question:  the scattering of the Earth and the Moon (here ignoring the side issue  that they form a bound state) is certainly a problem with large energy, $E>>>M_D$, and where the Einstein action makes excellent predictions -- how can this be?  A reason for this is the weakening of gravity with distance, which compensates for the large energy.  Likewise, we expect that in the relativistic regime, we can learn more if we allow ourselves to specify the impact parameter $b$, which is a measure of the distance scale probed, along with $E$.  A schematic plot of the different scattering regimes in the $E-b$ plane is shown in fig.~\ref{Ebdiag}.  Let's discuss some of its features. 

First, as we raise the energy, we can probe to shorter distances, staying above a line determined by the uncertainty principle.  This is the story of high-energy accelerators probing smaller scales, and followed to its end reaches the uncontrolled, nonrenormalizable Planck regime, $E\sim M_D$, $b\sim 1/M_D$, and marked ``NR".  But, suppose we control the impact parameter -- like one really does in accelerators -- and try to exceed $E=M_D$.  For large enough impact parameter, there is no obvious obstacle to still using the Born approximation. 

Next, let us investigate the scattering behavior in such a  ``phase diagram" by fixing the energy $E\gg M_D$ and lowering the impact parameter.  The Born approximation ultimately does fail, due to the contribution of higher-loop diagrams, but those of a very specific kind -- the ladder diagrams, as pictured, which add up to give an eikonal approximation to the amplitude, closely related to the classical approximation.  Earth-Moon scattering in fact lies in this eikonal regime, which we will study shortly.  

But first, what happens for smaller impact parameters?  While not all details are worked out, and ultimately a profound puzzle remains, an apparently consistent picture is the following. The scattered objects are deflected in each other's gravitational fields, and thus radiate.  Moreover, if they are composite ({\it e.g.} hydrogen atoms, or strings), their internal degrees of freedom can become excited once tidal gravitational forces pass a threshhold.  Thus the picture changes in a model-dependent way at some impact parameter depending on which ``internal" dynamics becomes important first; this is  indicated by the dotted line.

However, with present knowledge it appears that this kind of model-dependent behavior doesn't interfere with consideration of even smaller impact parameters, where something apparently very model-independent happens.  Namely, once the impact parameter reaches the scale of the Schwarzschild radius corresponding to $E$, gravity becomes strong, and classically a black hole would form\cite{EaGi}.  Our ultimate and central question is how to give a quantum description of this scattering regime.

Given the nonrenormalizability of gravity, mention of the word ``loops" may immediately ring an alarm bell -- how can we trust any of the story, once they become important?  But, despite these concerns, our approach to the problem seems to make sense.  First off, nonrenormalizability and the corresponding large contributions of higher-dimension operators appear to be {\it short distance} issues, and we are describing phenomena at large, even macroscopic, distances.  Moreover, we can investigate this story rather explicitly, using various candidate regulators for the theory.  One would be a na\"\i ve cutoff on loop momentum, but more sophisticated cutoff prescriptions are provided by supergravity (at least for sufficiently low loop level), or string theory!  I will summarize some of the things we can say so far, and give a picture that while provisional, appears to hang together well.

\subsection{Eikonal regime, and classical scattering}

\begin{figure}[!hbtp]
\begin{center}
\includegraphics[width=8cm]{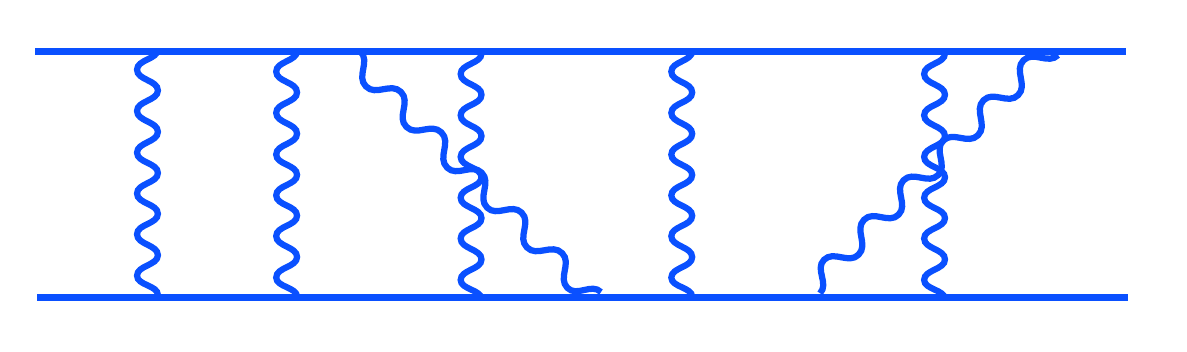}
\caption{A ladder diagram, with multiple graviton exchange.  (Note that the crossed lines do not connect.)}\label{fig:eik}
\end{center}
\end{figure}

I have claimed that ladder diagrams (together with crossed ladder diagrams) are the first loop diagrams to become important as the impact parameter is lowered; let's investigate them more closely.  One is shown in fig.~\ref{fig:eik}.  The nice thing is that they can be approximately computed, and summed up, for small enough momentum transfer $q=\sqrt{-t}$.  

\begin{figure}[!hbtp]
\begin{center}
\includegraphics[width=8cm]{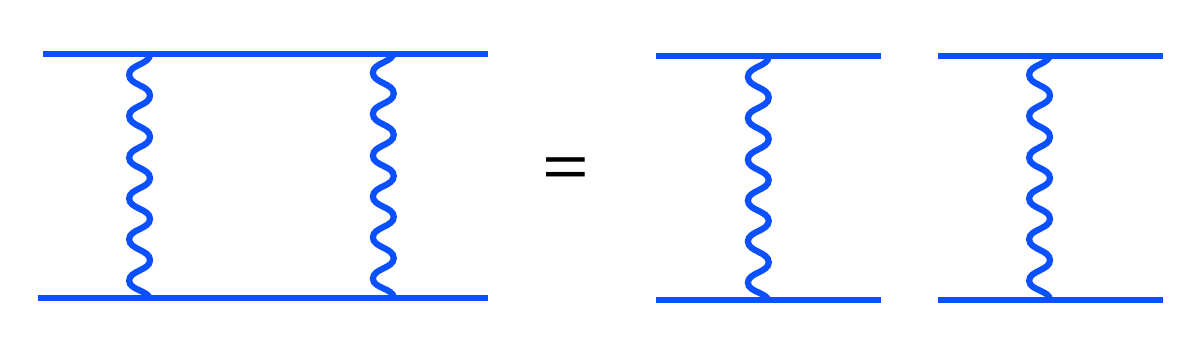}
\caption{The one-loop ladder diagram is obtained by combining two tree diagrams.}\label{fig:oneloop}
\end{center}
\end{figure} 

Consider for example the one-loop ladder.  As shown in fig.~\ref{fig:oneloop}, this can be thought of as two tree diagrams, sewn together.  Specifically, if the momentum transfers $k_i$ by the individual gravitons are small, we can replace the scalar propagators using
\beq\label{approxprop}
{-i\over (p-k_i)^2+m^2 -i\epsilon} \approx {i\over 2p\cdot k_i +i\epsilon}\ .
\eeq
Adding the crossed diagram, it is a straightforward exercise to compute the combined one-loop amplitude.  This is most easily written in terms of the Fourier transform of the tree amplitude, with respect to the momentum transfer $q_\perp$ perpindicular to the CM momentum,
\beqa\label{eikphase}
\chi(b,s) &= {1\over 2s}\int{d^{D-2}q_\perp\over(2\pi)^{D-2}}
\, e^{i{\bf q}_\perp \cdot \bv}T_{\rm tree}(s,-q^2_\perp)\cr &= {4\pi\over (D-4)\Omega_{D-3}}  {G_D s\over b^{D-4}}\ ,
\eeqa
where 
\beq
\Omega_n={2\pi^{(n+1)/2}\over \Gamma\left[(n+1)/2\right]}
\eeq
is the volume of the unit $n$-sphere.
We refer to $\chi$ as the {\it eikonal phase}. In the first line, we could have considered the tree-level amplitude for some other theory, for example string theory.  The second line is the result using the Born amplitude for Einstein's action, (\ref{bornamp}), but for small momentum transfers string or other corrections to this would be very small anyways.  The variable $\bv$ is a vector-version of the impact parameter.  The combined one-loop amplitude has a very simple form in terms of $\chi$:
\beq
T_1\approx 2s \int d^{D-2}b\, e^{-i{\bf q}_\perp\cdot \bv} {[i\chi(b,s)]^2\over 2!}\ .
\eeq

\ex{By combining the one-loop ladder and crossed-ladder diagrams, and using the approximation (\ref{approxprop}), derive this equation.}

Once you've seen the derivation of this, and its simple form, generalization to the $N$-loop ladder diagram is fairly clear.  The powers of $i\chi$ give an exponential sum, and the full {\it eikonal amplitude} is
\beq\label{eikamp}
iT_{\rm eik}(s,t) = 2s \int d^{D-2} b\, e^{-i{\bf q}_\perp \cdot \bv}[e^{i\chi(b,s)} -1]\ .
\eeq
And this now shows the origin of the top line on the right side of fig.~\ref{Ebdiag}.  For $\chi\ll 1$, the sum is approximated by the linear term, which  is just the Born amplitude.  For $\chi\roughly>1$, one must sum higher terms in the series.  Indeed, in the latter regime, the Born amplitude becomes large; the full sum then {\it unitarizes} the amplitude, as can be seen from the partial-wave expansion\cite{GiSr}.  The crossover line given by $\chi\sim 1$ corresponds to 
\beq\label{eikimp}
b\sim (G_D E^2)^{1/(D-4)}\ ,
\eeq
as follows from (\ref{eikphase}).

But now, we encounter the question of short distances.  Expanding out (\ref{eikamp}) in powers of $\chi$ yields terms that are increasingly badly divergent as $\bv\rightarrow0$ -- in fact, the approximation (\ref{approxprop}) has worsened the bad short-distance behavior responsible for nonrenormalizability.  From this viewpoint, the eikonal amplitude (\ref{eikamp}) seems nonsensical.  Fortunately, this is not a good viewpoint.

To see this, consider a simplified version of (\ref{eikamp}),
\beq
I=\int_{M^{-1}}^1db b^3 e^{ig/b^2}\ .
\eeq
This looks a lot like (\ref{eikamp}) for $D=6$, but the exponential of $q_\perp \cdot \bv$, which makes the integral finite at large $b$, has been replaced by an upper limit, and an explicit cutoff $1/M$ has been introduced to regulate the short-distance behavior.  Expanding the exponential gives
\beq
I(M)={1-M^{-4}\over 4} + ig{(1-M^{-2})\over 2} - {g^2\over 2} \log M + {ig^3\over 12}(1-M^2)+\cdots\ ,
\eeq
which is terribly divergent as $M\rightarrow\infty$.  {\it But}, plugging into Mathematica (or using a table of integrals, for the old-fashioned) gives
\beq
I(M) = {g^2\over 4} \left[Ei(ig) +{1\over g}\left(i+{1\over g}\right)e^{ig} - Ei(igM^2) -{1\over gM^2}\left(i+{1\over gM^2}\right)e^{igM^2}\right]\ ,
\eeq
with $Ei$ the exponential integral,
and this approaches a finite limit, with controllable small corrections, as $M\rightarrow\infty$!  

So, expanding the integral (\ref{eikamp}) was not a good thing to do, and this carries  an important lesson.  Instead, we can see that (\ref{eikamp}) has a saddlepoint at 
\beq\label{saddleq}
q_\perp\sim \partial \chi/\partial b\ , 
\eeq
and the saddlepoint approximation is a good way to approximate the integral.  For large $E$ and  $q/E\ll 1$, this saddlepoint is at {\it long distances}
\beq\label{b-eik}
b\sim [G_D E (E/q)]^{1/(D-3)}
\eeq
-- so the short distance behavior doesn't play a role, and likewise short distance corrections to the integrand are not expected to be significant.  Apparently, {\it short distance dynamics is essentially irrelevant}.  Moreover, for a good approximation to the amplitude, we clearly need the full behavior of the summed exponential.

This sum introduces a connection with classical scattering.\footnote{For some early discussion of this connection, see \cite{tHoo,ACV}.}
Consider the classical metric of a particle of mass $m$ that has been boosted to very high energy, $E/m = \gamma\gg 1$, but take $m$ to zero such that $E$ is fixed.  The resulting metric is most easily explored in light-cone coordinates, $x^\pm=t\pm z$, $x_\perp$, where it takes the form found by Aichelburg and Sexl\cite{AiSe}:
\beq\label{ASmet}
ds^2 = -dx^+dx^- + dx_\perp^2 + \Phi(x_\perp) \delta(x^-) dx^{-2}\ ,
\eeq
with
\beq
\Phi= -8 G_D E\log(x_\perp)\ ,\ D=4\quad ;\quad \Phi= {16\pi G_D E\over \Omega_{D-3}(D-4)x_\perp^{D-4}}\ ,\ D>4\ .
\eeq

\ex{Derive this metric, by considering the stated limit of the Schwarzschild solution \eqref{Dschwarz}.  Hint: work in isotropic coordinates.}

The departure from a flat metric is completely supported in a shock wave at $x^-=0$; like with a boosted charge, the field lines pancake up into the transverse plane.  Despite the delta function in the metric, one may solve the geodesic equation for a particle incident along $-z$ that scatters at impact parameter $b$, and derive a finite scattering angle.  One finds that this scattering angle matches the momentum transfer given by the eikonal saddlepoint; note also $\chi \propto E\Phi(b)$.

\ex{Check these statements.}

Returning to the question of short vs. long distance physics, one can understand why even very ultraplanckian scattering is only probing long-distance physics, in terms of a phenomenon we will call ``momentum fractionation\cite{GSA}."  For a given momentum transfer $q$, which can also be ultraplanckian, we expect the scattering to be dominated by impact parameters near that  given by \eqref{b-eik}; if the corresponding eikonal phase $\chi$ is large, then from the exponential in (\ref{eikamp}) we see that the dominant loop order in the sum is $N\sim \chi$.  Then, in the ladder diagram fig.~\ref{fig:eik}, one estimates a typical momentum $k\sim q/N\sim (\partial \chi/\partial b)/\chi$ to flow through each rung.  This gives $k\sim 1/b$, which is very small as long as $b\gg M_D$.  Large momentum transfer fractionates into many small transfers, and one effectively only has a soft probe of the dynamics.  

\subsection{Match to supergravity amplitudes}

\begin{figure}[!hbtp]
\begin{center}
\includegraphics[width=15cm]{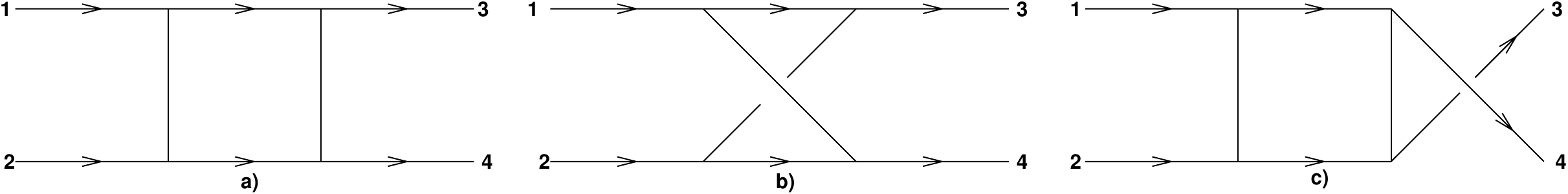}
\caption{All one-loop, four-point supergravity diagrams can be written in terms of the three scalar box diagrams shown.}\label{fig:sugraone}
\end{center}
\end{figure}

If one still harbors suspicion about such cavalier treatment of divergent loop amplitudes, these can be examined in a case where they are manifestly convergent -- for low enough loop level (or possibly to all order \cite{BDR}), in supergravity.  This was done in \cite{GSA}.  In particular, the explicit, finite one loop amplitudes given in \cite{BDDPR} take the form (up to polarization dependence)
\beq\label{sugone}
M_1(s,t) = -i (8 \pi  G_D)^2
                  s^4  
                  \left[ I^{\rm 1}(s,t)+I^{\rm 1}(t,u)+I^{\rm 1}(s,u)\right]\ ,
\eeq
given in terms of the integral
\beq
I^{\rm 1}(s,t) = \int {d^D k \over (2\pi)^D}
                        { 1 \over k^2 (p_1-k)^2 (p_2+k)^2 (p_1+p_3-k)^2 }\ .
\eeq
For non-supersymmetric gravity, the numerator in the one-loop integrals would have additional terms $\sim (k/E)^p$, giving divergent behavior -- thus, we see that the effective cutoff on $k$ provided by supergravity is at $k\sim \sqrt s$.  
Interestingly, the expression (\ref{sugone}) captures {\it all} of the complicated one loop structure, in a small set of diagrams easily matched with those of the eikonal approximation!  These are shown in fig.~\ref{fig:sugraone}; the first two are clearly identifiable with the ladder and crossed ladder diagrams.  The third is suppressed in an expansion in $q^2/E^2\sim -t/s$, as it requires a large momentum exchange through a single line.  Examining the explicit expressions, we see that 
\beq
M_1(s,t) = T_{eik}^{1\,loop} + {\cal O}(t/s) + {\rm cutoff\ dependent}\ .
\eeq

\begin{figure}[!hbtp]
\begin{center}
\includegraphics[width=12cm]{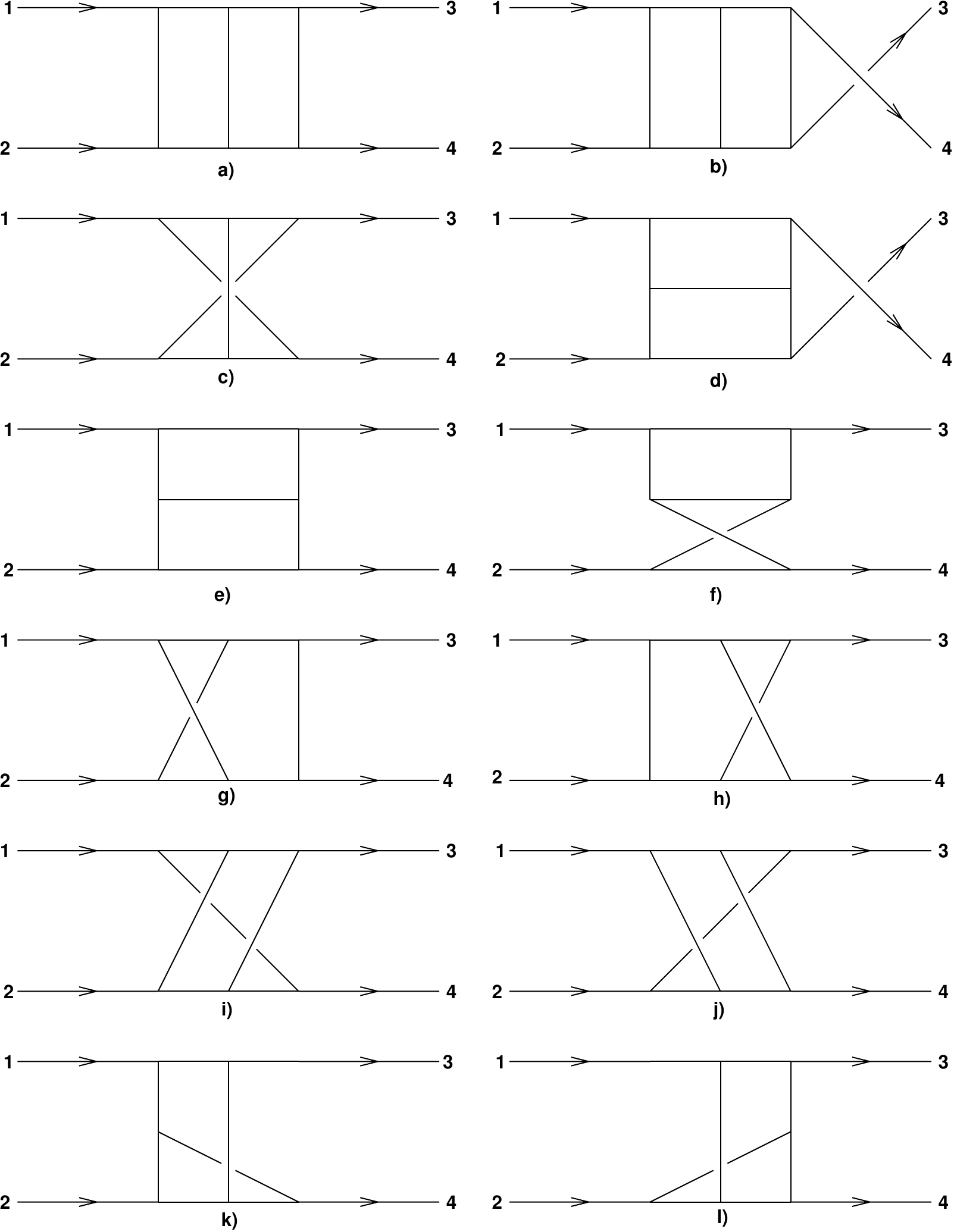}
\caption{All two-loop, four-point supergravity diagrams can be written in terms of the scalar diagrams shown.}\label{fig:sugratwo}
\end{center}
\end{figure}

Going further, the two-loop supergravity diagrams are also finite.  We again find\cite{GSA}, using the explicit SUGRA results of \cite{BDDPR},  
\beq
M_2^{SUGRA}(s,t) =  T_{eik}^{2\,loop} +{\cal O}(t/s) + {\rm cutoff\ dependent}\ .
\eeq
The two-loop diagrams of \cite{BDDPR} are shown in fig.~\ref{fig:sugratwo}; those in a), c), g), h), i), and j) correspond to the ladder and crossed ladder diagrams, and the rest are subleading in $t/s$.

This story illustrates another important phenomenon, that of {\it graviton dominance}\cite{tHoo,GSA}.  Specifically, all the states of the supergravity multiplet contribute to the amplitudes, yet we have just found that the gravitational exchange dominates at high-energies and long distances.  This is generic, and can be explained\cite{GSA} by noting that the coupling of an exchanged particle with a given spin grows like 
\beq
E^{\rm spin}\ ,
\eeq
basically from couplings to $\partial_\mu^{\rm spin}$.  

\ex{Explain this more carefully, investigating explicit couplings.}  

Thus, the high-energy, long-distance behavior is expected to be relatively generic to any theory of gravity.  One caveat is that the emitted radiation will depend on the spectrum of light states -- though this is expected to be a subleading effect\cite{GiSr}.

\subsection{Momentum fractionation and strong gravity}

Returning to momentum fractionation, we see it has some apparently important consequences\cite{GSA}.  First, it apparently presents a significant challenge to a meaningful formulation of the proposed phenomenon of {\it asymptotic safety}\cite{Wein, Niedermaier:2006wt} in terms of physical amplitudes\cite{GSA,Dvali:2010ue}.  Specifically, if one wished to find a physical probe of this phenomenon, which is basically running of the gravitational coupling to a non-perturbative fixed-point at large energy, ultraplanckian scattering is the obvious place to look.  But, the preceding arguments indicate that the gravitational coupling constant only enters the ultrahigh-energy scattering at low momentum transfer; for example, the argument involving eikonal diagrams shows that they depend on $G_D(q)$ at momentum transfer $ q\sim k\sim 1/b$, and are not sensitive to Newton's constant at $q\gg M_D$.  The same logic obstructs a meaningful probe of hypothesized Regge behavior in gravity.

This UV/IR behavior, telling us that in high-energy gravitational scattering we are really probing long distances, appears important and profound.  In particular, we have encountered the problem of nonrenormalizability in  other incomplete theories, like the four-Fermi theory of weak interactions.  In these previous examples, one could also see the trouble with the theory through a breakdown of unitarity, and these problems are linked, as one finds by considering growth of amplitudes near the cutoff energy.  But, in gravity we encounter a unitarity problem of a very different character, apparently arising purely in {\it long distance} physics, and delinked from the issues of short-distance behavior, nonrenormalizability, and related considerations.

To begin to understand the origin of this problem, let us imagine lowering the impact parameter further, specifically to the value
\beq\label{Schb}
b\sim R(E)\sim (G_D E)^{1/(D-3)} .
\eeq
This is the parametric dependence of the D-dimensional Schwarzschild radius given by the center-of-mass energy, and corresponds to the line demarcating the strong-gravity region in fig.~\ref{Ebdiag}.  The former statement is seen explicitly from the form of the D-dimensional Schwarzschild solution,
\beq\label{Dschwarz}
ds^2 = -\left[1- {k_D  M\over M_D^{D-2}r^{D-3}}\right]dt^2 + {dr^2\over 1- {k_D M\over M_D^{D-2} r^{D-3}}} + r^2 d\Omega^2_{D-2}\ ,
\eeq
with $k_D$ a numerical constant.  At impact parameters below the value (\ref{Schb}), one classically expects a black hole to form.  

\begin{figure}[!hbtp]
\begin{center}
\includegraphics[width=4cm]{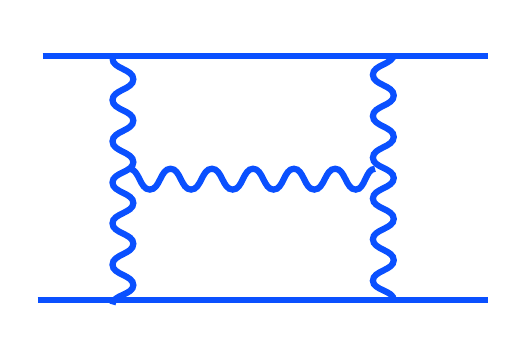}
\caption{The ``H-diagram" is one of a family of diagrams with graviton trees attached to the high-energy source particles; these produce subleading terms in an expansion in $t/s$.}\label{fig:H}
\end{center}
\end{figure} 

First, let us investigate this problem from the perturbative perspective.  Specifically, there are subleading loop diagrams we have so far neglected, like that shown in fig.~\ref{fig:H}.  These are subleading in an expansion in $  [R(E)/b]^{2(D-3)}\sim -t/s$, as can be seen from the Feynman rules.

\ex{Using the Feynman rules, infer that if one of the rungs of an eikonal diagram with typical momentum $k\sim 1/b$ is replaced with the H diagram of fig.~\ref{fig:H}, the resulting expression is indeed suppressed by a factor of $[R(E)/b]^{2(D-3)}$ relative to the original amplitude.}  

Indeed, we saw the presence of such subleading behavior both in the corrections to the eikonal amplitudes, and in the supergravity amplitudes.  These corrections are small as long as the scattering angle $\theta \sim \sqrt{-t/s}\sim [R(E)/b]^{(D-3)}$ is small.  But, they suggest that the perturbation series is no longer an expansion in a small parameter once $b\lesssim R(E)$, and in fact diverges.

A check on this comes from a simpler problem; in \cite{Duff}, Duff showed that an analogous sum over graviton tree diagrams sourced by a point mass produce the Schwarzschild metric.  The corresponding sum is indeed divergent at the Schwarzschild radius.  We likewise expect that the sum of tree diagrams attached to the high-energy sources gives the classical collision geometry, if we trust the picture to this point; there have been a variety of tests of this picture, though more are still being performed.  It was shown in \cite{EaGi} that this classical geometry contains a black hole.  Specifically, for $b\lesssim R(E)$, a closed-trapped surface, or apparent horizon forms; singularity theorems plus cosmic censorship then impliy a corresponding black hole.  

There are three further comments on this picture.  First, there will also be some classical radiation.  But, cosmic censorship gives a lower bound on the mass of the resulting black hole, since the horizon area can only grow from its initial value; such lower bounds are in the range of half the collision energy.  Second, as with Duff's simpler calculation, one expects a divergent perturbation expansion associated with black hole formation here:  a black hole is {\it not} a small perturbation of flat space.  Moreover, there is no indication of how features of string theory or supergravity will save us from this breakdown.  Now, we could take the resulting classical geometry as a new starting point, and try to quantize in a perturbation expansion about this.  The third comment is that this has been done, in cases with more symmetry, beginning with Hawking's classic demonstration of black hole evaporation\cite{Hawkrad}.  And, here we encounter the most serious problem: such a quantization led to the prediction of massive unitarity violation\cite{Hawkunc}.  

To summarize this section, at $E\gg M_D$ we can control the impact parameter by scattering wavepackets.  We can also see a correspondence with the semiclassical picture.  Classically, collisions at sufficiently small impact parameters produce black holes, plus some radiation.  Quantum corrections to this picture lead to Hawking radiation.  And, this leads to a unitarity crisis.  

To understand the last statement, we turn to a deeper examination of black hole evaporation.

\section{The gravitational unitarity crisis}

The situation I have presented leads to an apparently critical challenge to the foundations of present-day physics, through what has been called the black hole information paradox. I will give a lightening review of this; more details appear in other reviews such as \cite{Astrorev,SGinfo}.

As I have noted, Hawking found that perturbative quantization about a black hole geometry leads to its evaporation, with a
 final state that is mixed so that information is destroyed and unitary evolution fails\cite{Hawkunc}.  The analysis can be carried out even more explicitly in two-dimensional models \cite{CGHS,GiNe}.  This story represents, at the least, a breakdown of quantum mechanics, but we find that the situation is even worse.

\begin{figure}[!hbtp]
\begin{center}
\includegraphics[width=7cm]{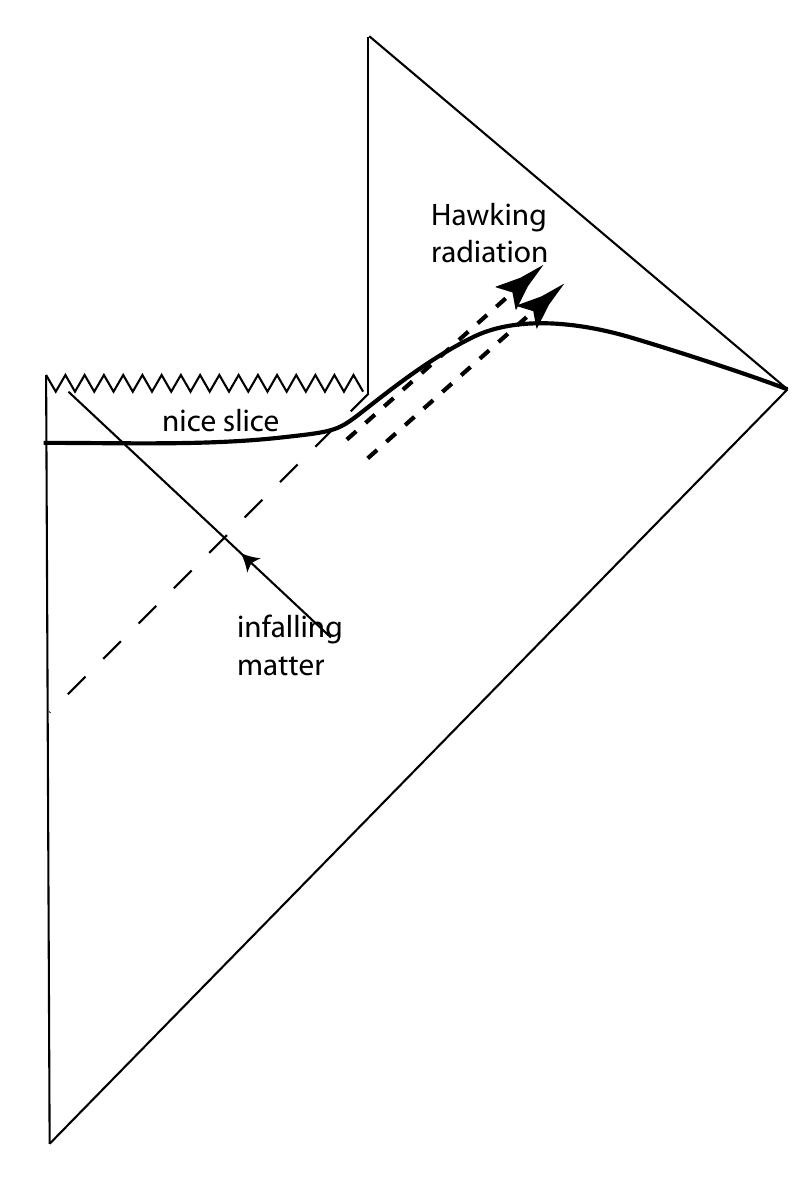}
\caption{A Penrose diagram of an evaporating black hole, with one of a family of nice slices pictured.}\label{Niceslice}
\end{center}
\end{figure} 

Let's start by giving an updated version of Hawking's argument -- the {\it nice slice} argument.  
First, a family of spatial slices can be drawn through the black hole, avoiding the strong curvature near the classical singularity, crossing the horizon, and asymptoting to a constant-time slice in the asymptotic flat geometry.  One such slice is sketched in fig.~\ref{Niceslice}. These ``nice slices" were described in \cite{LPSTU}, with one explicit construction in \cite{GiNLvC}.

Locality in field theory tells us that the state on such a slice can be represented as a sum of products of states from two Hilbert spaces, one corresponding to inside the black hole, and one outside.  Schematically, this can be written (for more detail, see {\it e.g.} \cite{GiNe})
\beq\label{nicestate}
|\psi_{NS}\rangle \sim \sum_i p_i |i\rangle_{in} |i\rangle_{out}\ .
\eeq
A description of the state outside the horizon is given by the density matrix formed by tracing over the inside Hilbert space:
\beq\label{rhohr}
\rho_{HR}\sim {\rm Tr}_{in} |\psi_{NS}\rangle\langle \psi_{NS}|\ .
\eeq
This is manifestly a mixed state.  In fact, one can trace over all degrees of freedom that have not left the vicinity of the black hole by a given time; {\it e.g.} one can describe the density matrix on future null infinity as a function of the corresponding retarded time $x^-$.  This density matrix has an entropy $S(x^-)= -{\rm Tr}[\rho(x^-)\log\rho(x^-)]$, and one straightforwardly estimates that it grows with $x^-$ to a value of order the Bekenstein-Hawking entropy, proportional to the original horizon area, $S(E)\propto G_D[R(E)]^{D-2}$, at a retarded time corresponding to the time the black hole evaporates to close to the Planck size.  This represents a huge missing information, which is also largely independent of what was thrown into the black hole.  This gives a modern update of Hawking's original argument\cite{Hawkunc} for the breakdown of unitarity.

The problem with the story is that quantum mechanics is remarkably robust.  Hawking proposed a linear evolution law for density matrices
\beq
\rho \rightarrow \slashed{S}\rho\ ,
\eeq
generalizing the S-matrix.  But Banks, Peskin, and Susskind\cite{BPS} argued that this causes severe problems with energy conservation.  A basic argument is that information transfer and loss requires energy loss, and that once allowed, energy non-conservation will pollute all of physics through virtual effects.  In particular, \cite{BPS} leads to the conclusion that the world would appear to be thermal at a temperature $T\sim M_D$.

If locality tells us that there is no information escape during Hawking evaporation, and energy conservation or quantum mechanics tell us that information is conserved, 
the obvious alternative is that information does not escape the black hole until it has reached the Planck size, where the nice-slice argument manifestly fails.  However, with an available decay energy $\sim M_D$ and an information $\Delta I\sim S(E)$ to transmit to restore purity, on very general grounds this must take a very long time, $\sim S^2$.  This implies very long-lived, or perhaps stable, remnants, which come in $\calo(\exp S(E))$ species ({\it i.e.} internal states) to encode the large information content.

Since one can in principle consider an arbitrarily large initial black hole, the number of species is unboundedly large.  This leads to unboundedly  large inclusive pair production in generic processes with total available energy $E\gg M_D$, as well as problems with inconsistent renormalization of the Planck mass\cite{Susstrouble}, {\it etc.}  (The former problem can be seen particularly clearly in the charged black hole sector\cite{wabhip}, where one considers Schwinger production.)  

So, information apparently cannot get out of a black hole, cannot be lost, and cannot be left behind, and this is the essence of the ``paradox."  It represents a deep conflict between basic principles of Lorentz/diffeomorphism invariance (on a macroscopic scale), quantum mechanics, and locality (also on a macroscopic scale).  These are the foundation stones of quantum field theory, which therefore must apparently fall.

Both quantum mechanics and Lorentz invariance appear very robust; simple modifications of them lead to violent contradiction with experiment.  On the other hand, locality is a concept that, as we have seen, is not even easily formulated in quantum gravity.  So, it is natural to propose that locality is not sharply defined in this context, and that this 
ultimately
underlies an explanation of how unitarity is restored.  If this is true, and black hole formation and evaporation is a unitary process, Page\cite{Page} has argued on general information-theoretic grounds that information must begin to be emitted by the time scale 
\beq\label{pagetime}
{t_{Page}\sim R(E)S(E)\ ,}
\eeq
 where the black hole has radiated an $\calo(1)$ fraction of its mass.  This indicates a needed breakdown of the nice-slice argument, and some departure from locality as described in the semiclassical picture of fig.~\ref{Niceslice},  over distances comparable to the black hole size, $\sim R(E)$ -- which can be a macroscopic scale. 

While other surprises in the context of high-energy gravitational scattering are not unfathomable -- we are still checking aspects of the picture outlined above -- it seems clear that black holes also form in collapse of massive bodies, yielding a version of the preceding argument for a conflict between basic principles.

\section{Nice slices and the local spacetime perspective}
\label{sec:localST}

The essence of the nice-slice argument can be summarized by:
\beq
|\psi\rangle\quad \rightarrow\quad  \rho={\rm Tr}|\psi\rangle\langle\psi|\quad \rightarrow\quad S=-{\rm Tr}\rho ln \rho =\Delta I\ .
\eeq
Specifically, evolution on the nice slices leads to a quantum state $|\psi\rangle$ describing the black hole plus surroundings.  Locality tells us this can be decomposed in terms of a product Hilbert space, representing distinct degrees of freedom inside and outside the black hole, and that moreover these degrees of freedom become entangled, and cannot become disentangled through escape of information from the black hole.  Finally, the von Neumann entropy $S$ gives a quantitative measure of the information that is ``stuck" inside the black hole.
Since this argument leads to an apparent paradox, and moreover it rests on the apparent ``weak link" of locality, it is worth examining more closely, to see if it has a loophole\cite{QBHB,GiNLvC}.  

To begin, note that the way the state $|\psi\rangle$ has been specified is gauge-dependent; it depends on some rather arbitrary, and in the case of the nice slices, extreme and artificial, choice of time slicing.  The question of a gauge-invariant specification encounters precisely the problem of gauge-invariant local observables outlined in section \ref{obssubsec}.  Specifically, we need to talk about information in localized degrees of freedom, either inside or outside the black hole, and so we need to describe this in terms of local observables, which we don't have.

Of course we know that local QFT has a built-in notion of local observables, and since our fundamental theory of gravity has a QFT approximation it should have an approximate notion of local observables.  The question of how these emerge from a gauge-invariant description is still being investigated, but we can outline a picture.  This picture is given in terms of the (effective) field theory description of gravity, which we don't believe to be the fundamental description, so it is incomplete.  But, nonetheless, it seems to give a suggestive guide, and moreover exploring its limitations plausibly hints at outlines of the more fundamental theory.

The basic idea has been outlined: approximate local observables emerge  in {\it relation} to features of the state.  This is because gauge transformations tell us that there is no intrinsic notion of location, but one can imagine locating something with respect to another feature of the state -- such as, {\it e.g.}, planet Earth!   In fact, this is also how current treatments of inflationary theory handle the problem of diffeomorphism invariance, which implies arbitrariness in the time slicing:  one commonly computes the perturbations, which we ultimately observe, at the reheating time. This is defined in the same kind of relational approach, as the slice where the inflaton field takes a specific value where reheating takes place.

Ref.~\cite{GMH} investigated construction of such gauge-invariant relational observables, that approximately reproduce local QFT observables in certain states.  
An example of such a ``proto-local" observable is an operator of the form
\beq\label{protoloc}
{\cal O} = \int d^Dx\sqrt{-g}\ O(x) B(x)\ .
\eeq
Here, $O(x)$ could be the local operator that we wish to discuss, and $B(x)$ is another operator acting on some ``reference background" fields with respect to which we localize. Such an expression is diffeomorphism invariant.  If we are in a state where the operator $B(x)$ is strongly peaked in some small region, then the integral \eqref{protoloc} essentially gives the operator $O(x)$ localized to that region.  In this sense, locality is emergent, and only in an approximation.  This approximation can break down, when fluctuations of the reference background are important, and/or it has a large backreaction.  An explicit toy model realizing such ideas can be given by working in 1+1 -dimensional gravity\cite{GaGiobs}  (aka: ``on the string world sheet").   One way to think of the reference background is in terms of  inclusion of a measuring apparatus.  The necessity of including such a background appears in harmony with a viewpoint enunciated by Bohr:  ``no phenomenon is a real phenomenon until it is an observed phenomenon."

If one needs a reference background in order to give a proper gauge-invariant description of the state, one can ask what kind of reference background is needed to resolve the features of the nice slice state.  A simple estimate is the following.  Hawking quanta are emitted with a typical energy $\sim 1/R(E)$, and roughly one such quantum is emitted every time $\sim R(E)$.  These have partner excitations, represented in \eqref{nicestate}, that fall into the black hole and should be registered in the internal part of the nice-slice state.  Thus, a reference background needs a spatio-temporal resolution at least to scales $\sim R(E)$, over time scales \eqref{pagetime}.  This means the reference background carries a minimum energy $S(E)/R(E)\sim E$, indicating that this would have a large backreaction on the black hole.

One might ask if the approximately local quantities like \eqref{protoloc} characterizing the nice-slice state might be defined using the existing background of the black hole, and thus avoid the need to include other excitations.  If so, one might ignore the preceding argument, and just try to give a QFT description of the nice slice state, working in a perturbative expansion in $G_D$ about the semiclassical black hole geometry.  The leading correction to this geometry, due to the expectation value of the stress tensor, just causes the black hole to shrink, as in \cite{CGHS}, and one can then try to compute the quantum state on slices in this evaporating geometry.  Here, though, one encounters an apparent issue as well\cite{QBHB}.  If one tries to compute the state $|\psi\rangle$ perturbatively, fluctuations about the evaporating geometry ultimately appear to have a large backreaction on $|\psi\rangle$, leading to a breakdown of the perturbative expansion.  This also occurs on the time scale \eqref{pagetime}.  

Thus, a proposed resolution\cite{QBHB,GiNLvC} of the information paradox is the following.  
The arguments leading to the conclusion that information is lost are not sharp; the semiclassical/perturbative nice slice picture is not an accurate representation of the detailed quantum state of a black hole.  In particular, it is not clear that the arguments can be made in a gauge-invariant fashion, and moreover a na\" \i ve attempt to quantize on nice slices encounters backreaction indicating a breakdown of such a perturbative treatment.  This means that we have not given a controlled perturbative argument for information loss,  and indicates that if one wants to examine the ultimate fate of the information, one needs to do so within a nonperturbative framework.  

With this proposed resolution, the paradox becomes a problem, that of determining the correct {\it nonperturbative} mechanics governing the theory, and in particular unitarizing the description of an evaporating black hole.  This may well not have the usual notions of locality when described with respect to the semiclassical evaporating geometry.  Moreover, if locality breaks down, or is not a ``sharp" notion, this statement needs to be relevant over distance scales that are large, of size $\sim R(E)$, which can be macroscopic, in order for information to begin to escape a big evaporating black hole by the time \eqref{pagetime}.

We would clearly like to understand these issues better.  It is always useful to have a simplifying toy model, and fortuitously we have one with a number of parallel features:  inflationary cosmology.  This of course has its own intrinsic interest(!), but also is a simpler place to examine closely related questions regarding dynamical evolution of geometry, and the role of observables, since it has greater symmetry. Space does not permit covering developments in this area here, but some highlights can be mentioned.  First of all, some investigation of approximately local observables, and limitations to recovering precisely local observables, are described in \cite{SGMadS} (see also \cite{GMH}).  Moreover, ref.~\cite{QBHB} explores the parallel between quantization on nice slices in a black hole background, and quantization on spatial slices in an inflationary background.   There an apparently analagous problem with perturbation theory was noted,  due to backreaction of fluctuations, appearing on the timescale $R S$, where now $R$ and $S$ are the Hubble radius and corresponding entropy.\footnote{The black hole/inflation parallel has also been in particular pursued by \cite{Nimatalk,Arkanietal}, who also argue for a breakdown at times $\sim RS$, but based on apparently different considerations.}  More explicit calculations have been performed, and additional arguments provided for breakdown of a perturbative treatment at such times\cite{GiSlone,GiSltwo} (see also \cite{BHLS}), associated with growth of the effect of fluctuations.  It is clearly worthwhile to explore and sharpen this parallel further, particularly with an eye on the needed nonperturbative completion and its implications.

\section{Lessons from loops or strings?}

There are two particularly popular approaches to quantum gravity, string theory and loop quantum gravity: what do they say about these issues?

If we focus on the S-matrix, this is still largely problematic for loop quantum gravity, where people are working to recover a complete description of configurations closely-approximating flat space, and to give a description of scattering of perturbations about it.  A modest first goal for this program would be, for example, to rederive the Born and eikonal amplitudes.  A general concern is that the constructions of this program do appear to modify a local description of spacetime, but only at distances comparable to the Planck length, $\sim 1/M_D$.  But, as I have just argued, the unitarity crisis really seems to require modifications of locality that can be relevant at very long distance scales, in appropriate circumstances.

String theory initially seems more promising.  It has a mechanism for nonlocality: the extendedness of the string, which can become large at high energies.  String theory has also furnished us with perturbative calculations of the S-matrix that avoid the infinities I described in discussing nonrenormalizability.  And, there are candidate non-perturbative descriptions of string theory; the most promising have been argued to be via dualities such as AdS/CFT, to more complete theories such as matrix theories or ${\cal N}=4$ super Yang-Mills theory.  Finally, it has been argued that these dualities realize the idea of {\it holography}, that a $D$-dimensional gravity theory in a region has an equivalent description in terms of a $D-1$ -dimensional theory on the boundary of that region -- a notion that, if true, is clearly nonlocal.

In physics one must pose correct, sharp questions, and so we should carefully investigate what string theory actually says.  We will only overview some of the basic arguments; a more complete treatment would involve another set of lectures.

First, consider extendedness.  This was conjectured to either interfere with black hole formation\cite{manyPC} or to provide a mechanism for information to escape a black hole\cite{LPSTU,Venez04}.  A way to test this is to explore the behavior of strings gravitationally scattering in a high-energy collision.  In such a collision, a string can become excited, and thus extended.  This was initially described as ``diffractive excitation" in \cite{ACVCQG}.  There is a clear intuitive picture for the phenomenon\cite{LQGST}, in terms of one string scattering off the Aichelburg-Sexl metric \eqref{ASmet} of the other: the metric produces tidal forces, which excite the string.  There is a threshhold impact parameter below which this effect occurs, given by\cite{ACVCQG,LQGST} 
\beq
b_t\approx {1\over M_D} \left(\sqrt{\alpha' }E\right)^{2/(D-2)}\ ,
\eeq
where the tidal impulse is sufficient to excite the string -- this is represented by the dotted line in the eikonal region of fig.~\ref{Ebdiag}.
A similar effect is expected in scattering of other composite objects, {\it e.g.} a hydrogen atom or a proton, with different threshholds given in terms of the excitation energy.

\begin{figure}[!hbtp]
\begin{center}
\includegraphics[width=13cm]{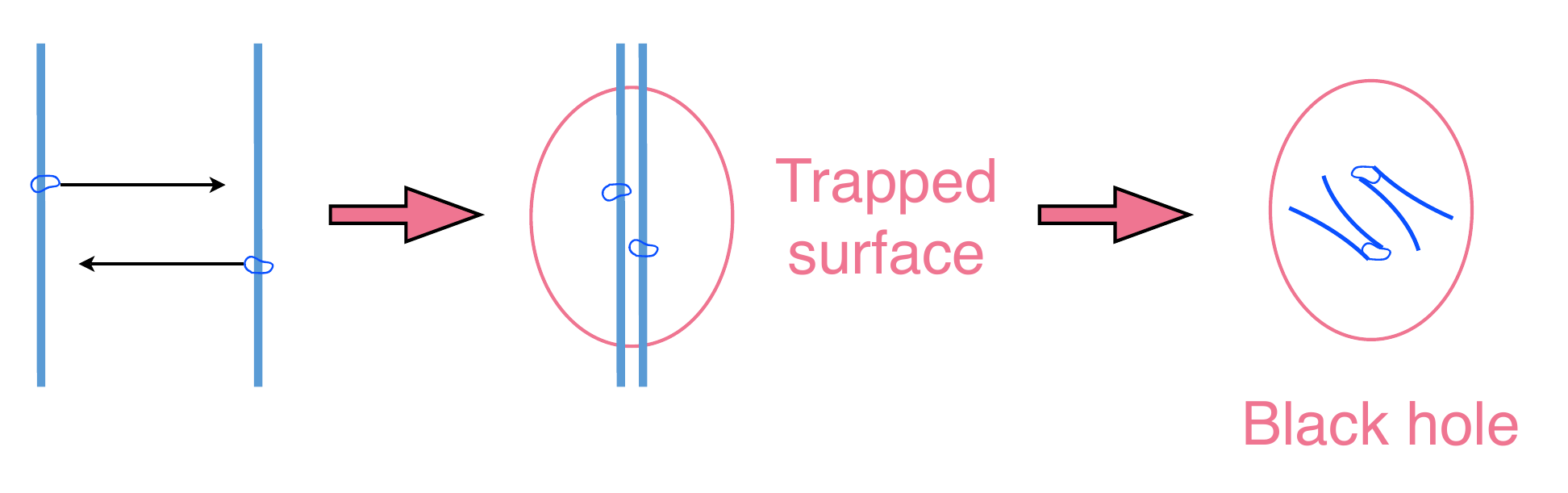}
\caption{Pictured is the scattering of two ultra-high energy strings; it is argued in the text that a trapped surface (hence black hole) forms before the strings spread out in response to the tidal impulse they receive from each others gravitational shock.}\label{fig:stringcoll}
\end{center}
\end{figure} 

String dynamics appears to respect notions of causality very similar to field theory.  In particular, if tidal excitation is what excites the colliding strings, one would expect that only to happen only after one string had met the Aichelburg-Sexl shockwave of the other\cite{LQGST}.  This can be tested\cite{GGM}, by explicit quantization of a string propagating in such a metric, where one indeed finds that the string spreads out after the collision, in response to the tidal impulse.  On the other hand, the trapped surface, thus black hole horizon, forms {\it before} the two shockwaves meet.  Thus, as illustrated in fig.~\ref{fig:stringcoll}, one ultimately excites the strings inside a black hole.  Then, there so far doesn't appear to be any good reason that the excited string can escape the black hole -- it is expected to behave like any extended object inside a black hole, and evolve locally/causally enough that it doesn't escape. There is thus no clear reason to believe that extendedness provides a mechanism for information to get out.

String theory is also famous for the improved properties of its perturbation theory, and in fact arguments for order-by-order finiteness in the loop expansion.  However, as I have described, this expansion appears to break down precisely in the circumstances where we encounter our unitarity crisis, at $b\sim R(E)$.  This statement, as explained at the end of section \ref{Pertquant}, appears generic to gravity, and there is no good indication that string theory escapes it; indeed, momentum fractionation indicates that the problem occurs independent of the short-distance degrees of freedom of the theory.  Similarly, conjectured perturbative finiteness of supergravity\cite{BDR} encounters the same limitation.  Again, it appears that unitarity is a more profound problem than short-distance issues such as renormalizability.

But, if we need a non-perturbative description of the theory to confront these issues, AdS/CFT and related dualities have been claimed to furnish one.  So, we should see if these solve our difficulties.  In order to isolate the effects of the curvature of anti de Sitter space from the strong-gravity dynamics of black holes, we would like to take the AdS radius to be large, $R\gg 1/M_D$, and see whether we can extract, in this regime, an approximation to the flat-space S-matrix.  Even if we can find a unitary S-matrix describing black hole formation and evaporation, we might not be satisfied that we have addressed the actual ``paradox," since it results from the conflict I have described between the description of localized degrees of freedom inside a black hole and a unitary description.  So far, in AdS/CFT, we do not have a description of localized observables, which we might expect to fit into our discussion of protolocal observables in the preceding section, so a local description of black hole interiors is lacking.\footnote{One idea is to seek an appropriate notion of observables that are relational in the matrix space.}  Nonetheless, we might justifiably feel that AdS/CFT providing such an S-matrix strongly indicates we're on the right track.

 There has been significant discussion of the problem of defining such an S-matrix purely from CFT data, without additional input from the bulk theory -- this is what is needed if the CFT is to give a non-perturbative definition of the theory.  Relevant work includes \cite{PolS,SusS,FSS,GGP,GaGiAdS,HPPS,katzetal,IsIt}.  
 
 In particular, a boundary CFT definition of the bulk theory would be achieved by a unitary equivalence between the CFT Hilbert space $\calh$ and a Hilbert space of bulk states.  If we consider a large AdS radius, say $R\sim 10^{10}$ lightyears, then this bulk Hilbert space should include the kinds of scattering states that we are familiar with, {\it e.g.} two particles in gaussian wavepackets, separated by 10 km, and prepared such that they will collide in a small region, resulting in outgoing states with similar asymptotic identification.  

In the case where the bulk string coupling constant $g_s\sim g_{YM}^2$ vanishes, also implying $G_D=0$, this can be achieved; free particle states of the bulk can be identified with states created by boundary operators, as reviewed in \cite{MAGOO}.  One might expect this holds also for small coupling.  However, we can think of AdS as providing a gravitational ``box" of size $R$, and the states of the free-particle Hilbert space correspond to particles that ricochet back and forth across this box for infinite time.  Thus, while we would like to isolate the S-matrix corresponding to a single scattering event, such states undergo infinitely many interactions.  In this sense the coupling does not necessarily have a small effect, and it is not clear we can extract from the interacting boundary Hilbert space the approximate S-matrix corresponding to the type of situation described in the preceding paragraph.  One proposal for deriving the corresponding reduced transition matrix elements appears in \cite{katzetal}, but it is also not clear that this approach in fact overcomes this ``multiple-collision" problem.

\begin{figure}[!hbtp]
\begin{center}
\includegraphics[width=6cm]{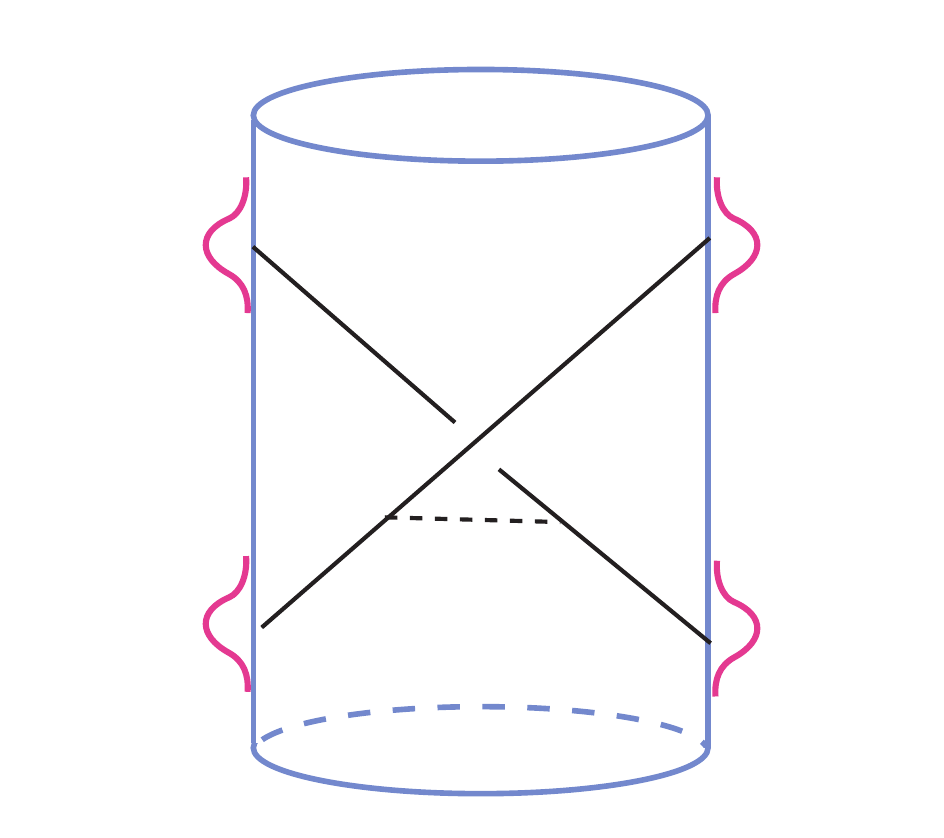}
\caption{Wavepackets created on the boundary of AdS scatter in the bulk.}\label{fig:adsscatt}
\end{center}
\end{figure}

In order to avoid this problem, Polchinski and Susskind\cite{PolS,SusS} proposed that one should instead consider states incident from the boundary of AdS, as pictured in fig.~\ref{fig:adsscatt}.  Here there are subtleties as well.  First, a boundary source generically produces a non-normalizable bulk state, that is thus not in the Hilbert space, and does not have a clear scattering interpretation.  A way around this is to prepare states using boundary sources with compact support, so that outside of this support the bulk state is normalizable.  

Indeed, \cite{GGP,GaGiAdS} explored such ``boundary-compact" wavepackets, and their scattering.  In particular, \cite{GGP} investigated the plane-wave limit of such wavepackets, and found some necessary conditions on the CFT if it is to produce familiar properties of the bulk S-matrix.  Specifically, the presence of a momentum-conserving delta function  requires a certain kind of singularity in correlators of the boundary theory, which had not been previously investigated, and behavior $\sim s^2/t$ in the Born regime requires a specific subleading structure to this singularity.  These necessary conditions appear non-trivial from the perspective of the boundary theory, although \cite{HPPS} provided some plausibility arguments that they are satisfied in certain types of CFTs, specifically those with large hierarchies of anomalous dimensions.   

However, \cite{GaGiAdS} pointed out another issue:  the boundary-compactness condition places limitations on the types of bulk wavepackets that can be obtained.  In careful treatments of scattering (see {\it e.g.} \cite{ReSi}) one uses wavepackets that fall at long distance faster than a power-law -- such as gaussians.  While boundary-compact wavepackets can be arranged to have characteristic widths, outside these widths they have tails that fall off as a power of the distance, $1/r^\Delta$, where $\Delta$ is the conformal dimension of the corresponding boundary operator.  This is an obstacle to approximating the usual space of scattering states.  Such tails can matter, when one is {\it e.g.} trying to resolve very small matrix elements.  For example, the matrix elements to individual quantum states in the black hole evaporation problem are expected to be of size $\exp\{-S(E)/2\}$, and these are thus expected to be obscured by such tails.

 A logical alternative to a complete equivalence between bulk and boundary theories, that allows one to derive all desired features of the bulk theory from the boundary, is that the boundary theory is a sort of effective theory for the bulk, that is, captures certain coarse-grained features, but not all the fine-grained detail of the bulk dynamics\cite{FSS,GaGiAdS}.  Certainly the powerful principles of universality and symmetry are at work, and provide at least part of the explanation for relations between bulk and boundary.
 
Complementarity is an idea that has been widely discussed, in association with holography, to provide a picture of how information escapes a black hole in Hawking radiation.  In the form advocated in \cite{STU}, this states that observations of observers who stay away from the black hole, and of those who fall in, are complementary in an analogous fashion to complementarity of observations of x and p in quantum mechanics, and thus should not simultaneously enter a physical description.   Picturesquely\cite{SuLi}, an astronaut (named ``Steve") falling into a black hole may feel healthy all the way to the vicinity of the singularity, but the external observer describes the astronaut as burning up at the horizon, in the process imprinting his bits of information on the Hawking radiation.  This seems to be a radical departure from local quantum field theory.  So far it has been hard to give a detailed account of this idea.  It may be that a significantly less radical notion of locality is all that is needed to resolve the problem \cite{GiNLvC}.  One possible distinguishing feature of different proposals is the thermalization or mixing time, which is the time scale on which information falling into the black hole undergoes quantum mixing with the degrees of freedom of the Hawking radiation.  Advocates of the complementarity picture\cite{SuLi,HaPr, SeSu} suggest this time scale is of size $R\log R$.  However, \cite{GiNLvC} suggests this time scale could be much longer; an upper bound is the time scale \eqref{pagetime} found by Page.  Such a longer time scale for mixing would represent a less radical departure from the semiclassical prediction \cite{Hawkrad} of infinite mixing time.

 Another proposed story for black hole evolution, arising from string theory and related to the preceding ideas, is the fuzzball proposal\cite{fuzz}.  This suggests that a macroscopic black hole is more fundamentally described in terms of an ensemble of ``microstates"  each of which significantly differs from the black hole geometry and from each other {\it outside} the horizon.  With such large variations in the geometry of the states of the ensemble, it is very hard to see how an observer falling into such a configuration would not be uncomfortably disrupted when falling through the would-be horizon.  Thus, such a proposal seems to fit into the general category of massive remnants\cite{BHMR}, where the black hole evolves into a configuration that is massive but no longer has a meaningful horizon -- it is more like a kind of exotic star.  An essential question with such a proposal is to understand how such a configuration would dynamically evolve from a matter configuration collapsing  to form a black hole, and to be consistent with our knowledge of black holes.  In particular, as  \cite{BHMR} outlined, after horizon formation this would require some nonlocal propogation of information -- of a specific type, for the fuzzball proposal.

 In short, the jury is still out on whether AdS/CFT, or other dualities, which offer a similar approach, can supply a definition of non-perturbative string S-matrix, sufficient to investigate our puzzles of gravitational scattering.  If it does, it is very mysterious how it does so, that is, how {\it all} the features of {\it approximately} local $D$-dimensional gravitational dynamics are encoded in a lower-dimensional theory.   In such a hypothesis, one would like to understand the mechanisms and principles both for how the boundary theory mimics this higher-dimensional physics of local QFT at a fine-grained level, and also for how it modifies the local QFT picture, particularly in describing the dynamics of black hole formation and evaporation.  Perhaps, via such dualities, string theory accomplishes such miracles -- or maybe we need to reach beyond.  Either way, it is important to understand the mechanisms and principles in operation. 

\section{Beyond local quantum field theory}

In reaching for the principles of a more complete theory, one can seek inspiration from what seems to have been a similar crisis encountered in physics:  that of the classical atom.  A classical treatment of the hydrogen atom yields singular evolution, as the electron spirals into the charge center at $r=0$, producing an ``atomic stability crisis."  Before knowing quantum mechanics (and in the absence of experimental data), one theoretical approach would be to modify the laws of physics near $r=0$ to smooth out this singular evolution.  But, the ultimate resolution was quite different; we discovered that there is a new scale, the Bohr radius -- indicated by experiment -- and that the laws of classical physics simply do not furnish a good description of the atomic electron at shorter distances.  Instead, they must be replaced by a new set of principles and a new framework in this domain; in particular, one encounters the uncertainty principle, and in the end describes the physics via the wave mechanics of quantum theory.  

Our present situation seems similar.  We have found a breakdown of the classical laws at the $r=0$ singularity of a black hole.  We might expect that there is some short-distance modification of the physics, say on the Planck length scale, $\sim 1/M_D$, that smooths out the singular evolution; perhaps it is provided by the short-distance regulation of the physics by loop quantum gravity or string theory.  But, the (experimental) fact that black holes exist together with the black hole information problem strongly indicate that such a short distance solution is not sufficient.  Instead, it appears that some new physical principles and framework beyond local QFT are needed, and that those should significantly modify the physical description on distance scales extending at least to the horizon radius, which can be a macroscopic length.  Thus, the information problem could be an important guide, just as was the stability problem of the atom.

If we want to understand the physics involved, there is a basic set of starting questions.  First, we would like to more sharply characterize the circumstances in which local QFT fails.  This is important, because it tells us where the {\it correspondence boundary} lies, where familiar constructs must match onto those of a more basic theory.  In the case of quantum mechanics, the correspondence boundary is described by the uncertainty principle and related statements, and this is a very important clue to the underlying physics.  A second question is what mechanisms are used to describe the dynamics -- for example what picture do we have of flow of information, relation to spacetime, {\it etc.}  A third, related, question is what physical and mathematical framework replaces QFT, and, how do the familiar properties of local QFT emerge from it in domains away from the correspondence boundary?  In trying to answer these questions, we might expect that unitarity in black hole evolution could be a key guide, along with the need to provide approximately local and quantum-mechanical descriptions for dynamical spacetimes such as cosmologies or black hole interiors.

We have specifically seen that there appear to be very good reasons to question locality, as described with respect to semiclassical geometry, at macroscopic length scales, under appropriate circumstances.  One reason for this is that in such a semiclassical description of the black hole geometry, in order to restore unitary evolution, it appears that information must be relayed from deep inside the black hole, near the singularity, to outside the horizon, over spacelike separations that can be large, $\sim R(E)>>> 1/M_D$.\footnote{This suggests some basic inaccuracy of this semiclassical picture.}  Also,  we see a gravitational growth in the effective size of objects, through scattering.  This is indicated, for example, by the scattering angle 
\beq
\theta\sim\left[ {R(E)\over b}\right]^{D-3} 
\eeq
for a particle scattering from another particle of energy $E$, and becomes large at impact parameters $b\sim R(E)$ growing with the energy, as I have described.  Finally, there is the lack of local observables; specifically, the construction of approximately local observables outlined in section \ref{sec:localST} fails\cite{GMH} at increasing distances when the energies involved are increased.

We in particular would like to more sharply describe the correspondence boundary where local QFT ceases to be a good description.  There have been a number of such previous proposals.  For example, it has been widely believed that a breakdown occurs when spacetime curvatures reach the Planck scale.  Or in string theory, a ``string uncertainty principle" has been formulated\cite{VenUP,GroUP}, augmenting the usual uncertainty principle by a term accounting for the growth of strings with energy,
\beq\label{STUP}
\Delta X \geq {1\over \Delta p} + \alpha' \Delta p\ .
\eeq
An alternative viewpoint, related to modified dispersion relations, is that validity of familiar QFT requires small momenta, $p< M_D$.  These are all statements relevant to the description of a single-particle state.  Holographic ideas have been encoded in a statement about multiparticle states, that local QFT is only valid in cases where one restricts the information content of a region, bounded by area $A$, as
\beq
I\leq A/4G_D\ ;
\eeq
this finds its most sophisticated realization in the covariant entropy bound of \cite{Fischler:1998st,Bous-cov}.

But, our discussion of scattering suggests a different boundary.  In quantum mechanics, the uncertainy principle represents a limitation on the domain of validity of the classical dynamical description, given by phase space.  If we consider QFT in semiclassical spacetime, the dynamical description is given in terms of Fock space states in the background geometry.  In the example of a flat background, it is not clear how to formulate a single-particle bound, such as \eqref{STUP}, respecting Lorentz invariance; a shortest distance is not a Lorentz-invariant concept.  (We also have a ``good" classical geometry, the Aichelburg-Sexl metric, describing a highly-boosted particle.)  But, consider a two-particle state, with, {\it e.g}, minimum uncertainty wavepackets, with central positions $x,x'$ and momenta $p=\hbar k$, $p'=\hbar k'$:
\beq
\phi_{x,k} \phi_{x',k'} |0\rangle\ .
\eeq
In local QFT, we allow states with arbitrary $x,x'$ and $k,k'$.  But, in light of our discussion, we recognize that gravity becomes strong, and that we cannot trust such a Fock-space description, when the inequality
\beq\label{Locbd}
|x-x'|^{D-3} > G |k+k'|
\eeq
is violated, with $G\sim \hbar G_D$.  This is the proposed {\it locality bound} of \cite{LocBd}.  It also has generalizations for $N$-particle states\cite{LQGST}, and within the context of de Sitter space\cite{SGMadS}.  This kind of bound thus might play a role similar to that of the uncertainty principle, and  serve as an important principle giving clues to the more fundamental dynamics.

In seeking the mechanisms, principles, and mathematical framework of a more fundamental ``nonlocal mechanics" of gravity, there are other important guides.  In particular, nonlocality in a framework with a QFT approximation generically causes serious difficulty, and indeed inconsistency.  The reason is that, with Lorentz invariance, locality and causality are linked, so nonlocality ({\it e.g.} signaling outside the light cone) implies acausality (propagation backward in time), and acausality in turn leads to paradoxes such as the ``grandmother paradox," where one sends a message back in time instructing an evil accomplice to kill ones grandmother before ones mother is born.  Moreover, locality is observed to hold to an excellent approximation!
So very important questions are how usual locality can be absent as a fundamental property of the theory, yet  emerge in an approximate sense, and how this can happen without associated inconsistencies -- nonlocal needs to be nearly local, in the appropriate contexts.
 If this line of investigation is correct, these should be important guides, and are very interesting and challenging problems.	
One can begin to think about these by considering how locality is sharply described, in the QFT approximation.  A standard formulation, termed ``microcausality" is that local observables must commute outside the lightcone,
\beq\label{Opcomm}
[{ O}(x),{ O}(y)]=0\quad {\rm for} \quad (x-y)^2>0\ .
\eeq
As we have indicated, though, such observables are expected to only emerge in an approximation from protolocal observables of the more fundamental theory.  In that case, we would like to understand what features of that theory lead to approximate validity of  \eqref{Opcomm}.  

This story requires deeper understanding of relational observables, which would involve another set of lectures.  But, there is another way to probe locality in QFT, that returns us to a main topic of these lectures: in terms of properties of the S-matrix.   An important question is to understand how to distinguish S-matrices from local theories, and one important characteristic appears to be  {\it polynomial} behavior at large momenta.  A rough idea for why this is so is that interactions involving powers of derivatives are local, but nonpolynomial interactions, for example involving $\exp\{\partial^n\}$, are not.

\section{The gravitational S-matrix: exploring general properties}

I have argued that the S-matrix (or related inclusive probabilities, in $D=4$) should be a well-defined object in a quantum gravity theory with a Minkowski-space solution, and  should be a good approximate concept if there are solutions closely approximating Minkowski space.  One possible source of clues to the principles of a theory can be sought in the general structure of this S-matrix.  Indeed, as was realized in earlier studies of the S-matrix, {\it e.g.} in the context of strong interactions, the combined principles of unitarity, analyticity, and crossing are very powerful constraints.  Combining these with other properties of a theory can yield nontrivial information; for example, as we've noted, string theory was {\it discovered} by Veneziano finding an expression for the S-matrix with duality between $s$ and $t$ channels.  We might anticipate that we could gain important knowledge about quantum gravity by combining general properties of gravity with this powerful framework.  Given our preceding discussion, we would particularly like to understand how locality and causality enter the story.  This section will briefly summarize some preliminary investigation into these questions; for more discussion, see \cite{GiSr,GiPo}.

\subsection{Unitarity, analyticity, crossing, \ldots}\label{UAC}

In the S-matrix approach, unitarity, $S^\dagger S=1$, is a basic assumption.  Analyticity and crossing are somewhat more subtle in a theory of gravity, as compared to theories without massless particles.  Let us make the ``maximal analyticity hypothesis," that the only singularities are those dictated by unitarity.  For purposes of illustration, as above we will think about a theory with massive matter, {\it e.g.} a scalar, coupled to gravity.  But, the masslessness of the graviton leads to various important new behavior.

To illustrate, in a theory with a mass gap, the commutativity of local observables \eqref{Opcomm} implies\cite{GellMann:1954db,Martin:1965jj} that amplitudes are polynomially bounded, for example at fixed $t$ and large complex $s$.  This, in turn, allows one to derive a bound on the total cross section known as the {\it Froissart bound,}
\beq
\sigma< c (\log E)^{D-2}\ ,
\eeq
for some constant $c$.  Gravity manifestly violates this bound; for example, simply from strong gravity/``black hole formation" we find a cross-section
\beq\label{BHX}
\sigma_{BH}\sim E^{(D-2)/(D-3)}\ ,
\eeq
and we find an even larger cross section from \eqref{eikimp}, describing eikonal scattering, $\sigma_E\sim E^{2(D-2)/(D-4)}$.  We would like to understand what other properties necessarily change in a theory with gravity.\footnote{Ref.~\cite{Dvali:2010jz} suggests similar properties, particularly growth of cross sections, in other massless, nonrenormalizable theories.}

\begin{figure}[!hbtp]
\begin{center}
\includegraphics[width=15cm]{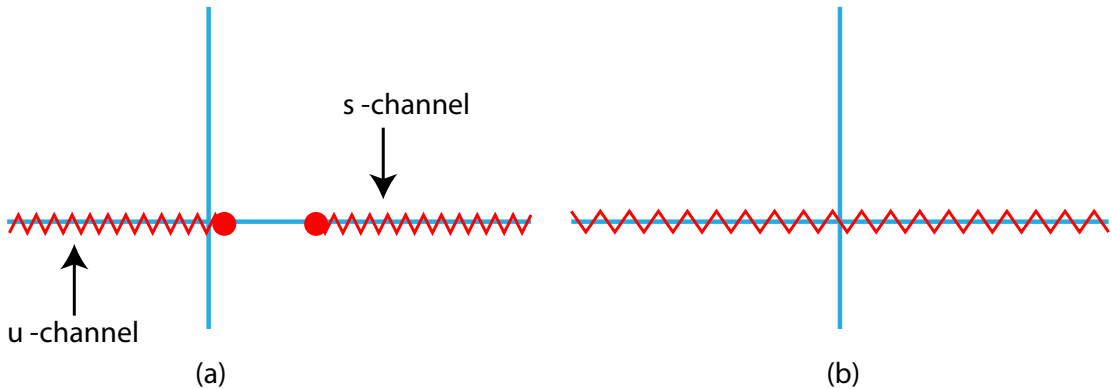}
\caption{(a) The complex s-plane in the case $m\neq0$.  (b) The complex s-plane in the case $m= 0$.}\label{fig:splane}
\end{center}
\end{figure} 

In fact, even crossing is more subtle in a theory without a gap.  Consider $2\rightarrow2$ scattering, and investigate the amplitude as a function of complex $s$ at fixed $t<0$.  In a massive theory we have cuts shown as in fig.~\ref{fig:splane}(a); the s-channel amplitude arises when we approach the right cut from above, but we can also analytically continue between the cuts and find the u-channel amplitude by approaching the left cut from below.  But, in a massless theory, the cuts merge, and our picture is instead fig.~\ref{fig:splane}(b).  Thus, we would seem to lose the ability to analytically continue between channels, which is an important property for constraining amplitudes.

So, clearly some properties familiar from massive theories are lost.  Without a theory that calculates the full gravitational S-matrix, we do not know which properties still hold.   One approach to further investigation is to explore which properties -- such as the preceding crossing path -- we are forced to give up, and which can be maintained:  in particular, we might assume that a familiar property from the massive case holds, until confronted with contrary information.  In this way, one can try to find the outline of a consistent story.  One set of checks on some of this discussion, which should be performed, comes from the fact that there is now significant explicit knowledge about supergravity to two and three loops, and beyond (see Bern's lectures at this school).  So, we might check which properties hold at least to a low, but non-trivial, loop order.

For example,  there is a different analytic continuation exhibiting  crossing, found by Bros, Epstein, and Glaser\cite{BEG}.  Begin by keeping $u<0$ fixed, and, at large $|s|$, follow the path $s\rightarrow e^{i\pi}s$.  Then, maintaining $s<0$ fixed, follow the path $t\rightarrow e^{-i\pi} t$.  This is an alternate path between the $s$ and $u$ channels.  (One also must check that the endpoints of the path can be connected.)

\ex{Check that the described path does indeed connect the $s$ and $u$ channels.  What assumptions are needed to connect the two path segments?}

\ex{Check whether this describes a valid crossing path at one, two, and three loops in $N=8$ supergravity, by working with the explicit amplitudes of that theory.}

Crossing symmetry can be combined with other properties to provide important constraints.  For example, another standard S-matrix property is hermitian analyticity, 
\beq
T(s^*,t^*) = T(s,t)^*\ .
\eeq

\ex{Determine whether the one, two, and three loop amplitudes in $N=8$ supergravity are hermitian analytic.}

\begin{figure}[!hbtp]
\begin{center}
\includegraphics[width=12cm]{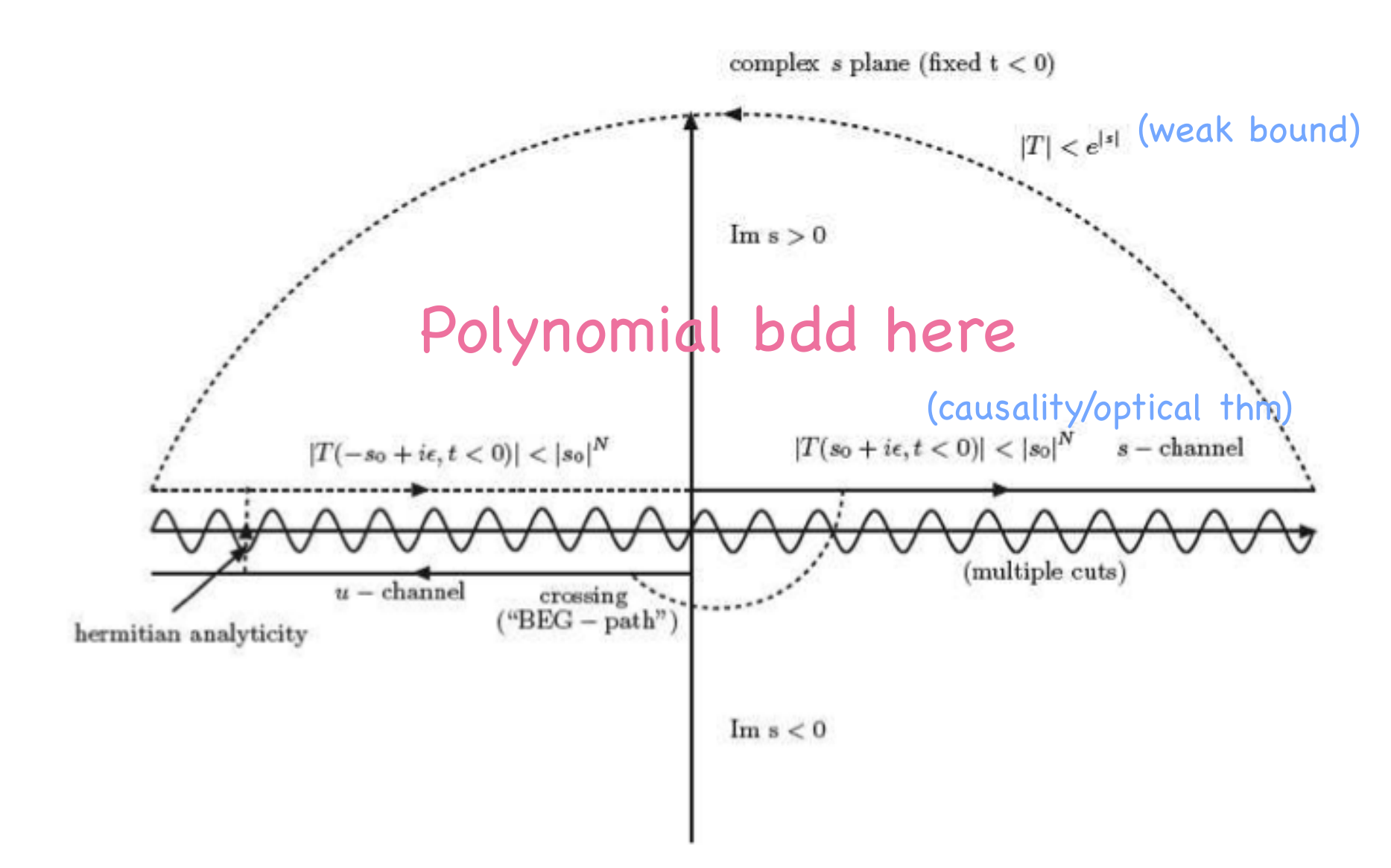}
\caption{Causality (a forward dispersion relation), crossing, and hermitian analyticity indicate a polynomially-bounded amplitude in the upper-half s-plane.}\label{fig:poly}
\end{center}
\end{figure} 

Crossing, hermitian analyticity, and causality potentially give information about polynomiality, which I have indicated is also connected with locality.  I will give a flavor of this, depending on certain assumptions that should be more carefully investigated.  Specifically, consider fig.~\ref{fig:poly}.  A basic statement of 
causality is that forward scattering ($t=0$) is polynomially bounded.  
With an additional assumption about the amplitudes being well behaved (this is an assumption of sufficient smoothness, bounding the real part of the amplitude, discussed in \cite{GiPo})
this can be converted into a statement that the amplitude is polynomially bounded as $s\rightarrow\infty$ along the upper side of the right cut. Crossing ensures the same behavior on the lower side of the left cut.  Then,  hermitian analyticity
implies the same bound on the upper side of the left cut.  Finally, if the amplitude satisfies a rather mild bound, $|T|<e^{|s|}$ in the upper-half s-plane, the  Phragmen-Lindelof theorem of complex analysis implies that in fact $T$ is {\it polynomially bounded} in the upper half plane\cite{GiPo}.  In this sense, causality plus apparently mild and plausibly true analyticity properties imply a polynomial bound in the physical region $t<0$ real, and $Im(s)>0$.  

However, this does not mean that the amplitude is {\it polynomial} -- it could behave nonpolynomially, but with nonpolynomial {\it growth} restricted to some other region besides the upper-half $s$ plane.  Indeed, basic features of gravitational scattering suggest that this occurs, in association with large cross sections such as  \eqref{BHX}. 

\subsection{Partial waves, strong gravity, and a ``Black-hole ansatz"}

One way to investigate the gravitational S-matrix is via the partial-wave expansion\cite{GiSr,GiPo}, which in $D$ dimensions takes the general form 
\beq\label{pwexp}
iT(s,t) = 2^{4\lambda+2}\pi^\lambda \Gamma(\lambda) E^{4-D} \sum_{l=0}^\infty (l+\lambda) C_l^\lambda(\cos \theta) \left[{e^{2i\delta_l(s) - 2 \beta_l(s)}}-1\right]\ ,
\eeq
where $\lambda = (D-3)/2$ and $C_l^\lambda$ is a Gegenbauer polynomial. Here $\delta_l(s)$, $\beta_l(s)$ are the real and imaginary (absorptive) parts of the phase shift for the $l$th partial wave, and the full $2\rightarrow2$ amplitude is encoded by them.    Basic features of gravity suggest specific behavior for these functions.  

For example, consider scattering in the strong gravity/``black hole" regime.  This corresponds to 
\beq\label{BHregion}
l\lesssim ER(E)\sim S(E)\ . 
\eeq
 We might expect that a quantum version of black hole states enter like resonances in two-body scattering.  Specifically, if such a black hole forms, it takes a time $\sim R(M)$ to decay to a different black hole state plus a Hawking particle, and thus has a width $\Gamma\sim 1/R(M)$.  Thus, such a state is very narrow: 
\beq
{\Gamma\over M}\sim {1\over R(M)M}\sim {1\over S(M)} \ll 1\ .
\eeq
For two-particle collisions with CM energy $E\sim M$ and angular momentum $l$, we expect a density of states of order
\beq
\rho_{acc}(M,l)\sim R(M)
\eeq
to be accessible directly via scattering\cite{GiPo}.
Their properties suggest properties of $\delta_l$, $\beta_l$.\footnote{One might ask about the full $\Delta{\cal N} \sim \exp\{S(M)\}R(M)\Delta M$ states in the interval $\Delta M$, expected from the Bekenstein-Hawking entropy $S(M)$.  Ref.~\cite{GiPo} argues that this larger set of states can be accessed by forming a black hole with mass $>M$, which then is allowed to evaporate to $M$.  This produces configurations where the $\Delta {\cal N}$ black hole states are entangled with the Hawking radiation emitted in reaching $M$.  For further discussion of properties of the spectrum and the connection with scattering, see \cite{GiPo}.}

For example, if we collide two particles at energy $E$ to form a black hole, and if Hawking evaporation at least gives a good coarse-grained picture, we expect there to be 
$\sim\exp\{S(E)\}$ outgoing multi-particle states.  (Here for simplicity we neglect the fact that $E\neq M$, which can be accounted for in a more careful treatment.)  Moreover, the amplitude for an outgoing state with just two particles is of size $\sim\exp\{-S(E)/2\}$.  The tininess of this corresponds to very strong absorption in the $2\rightarrow 2$ channel, and in partial waves can be parameterized by
\beq
\beta_l\approx {S(E,l)\over 4}\quad , \quad {\rm for} \quad l\lesssim E R(E)\ .
\eeq
This is a first part of a ``Black-hole ansatz" for the $2\rightarrow 2$ scattering\cite{GiSr}.  

We also expect features of the spectrum of states and scattering to be reflected in $\delta_l$.  In particular, in the case of a resonance with a single decay channel back to the initial state, Levinson's theorem tells us that increasing the energy through that of the resonance increases the phase shift by $i\pi$.  This suggests that accessible black hole states would lead to a formula $
\delta_l(E) \sim \pi S(E,l)$.  However, as discussed in \cite{GiPo}, the multiple decay channels lead to a more complicated story; we parameterize the second part of our ansatz as
\beq
\delta_l(E) = \pi k(E,l) S(E,l)
\eeq
where we expect $k(E,l)\sim {\cal O}(1)$ and $k(E,l)>0$, the latter condition corresponding to time {\it delay} in scattering.  

So, we have good indications of both strong absorption ($\beta_l$) and scattering ($\delta_l$) to large impact parameters, $l\sim ER(E)$, corresponding to the growing range of the strong gravity region, out to impact parameters $b\sim R(E)$. This na\"\i vely yields nonpolynomial $T(s,t)$, by \eqref{pwexp} producing exponential behavior $\sim \exp\{E^p\}$.  

A more complete story requires matching the physics onto the longer-distance region.  This produces an even larger cross section and range, as indicated {\it e.g.} by \eqref{eikimp}.  We also find corresponding nonpolynomial behavior in the eikonal amplitude  \eqref{eikamp}, as is seen from its saddlepoint approximation, which using \eqref{eikphase}, is
\beq\label{eikasymp}
T_{\rm eik}\sim\exp\left\{ i [s (-t)^{(D-4)/2} ]^{1/(D-3)}\right\}\ .
\eeq
This expression also exhibits nonpolynomial behavior.  

\ex{Derive \eqref{eikasymp} from the saddlepoint of \eqref{eikamp}.}

\subsection{Locality vs. causality}

An interesting question is whether quantum gravity can have an appropriate sense of causality (which I have argued is linked with consistency, at least in the semiclassical approximation), yet not be local.  The preceding discussion suggests a possible realization of these ideas, in terms of properties of the S-matrix.  

I have outlined in section \ref{UAC} an argument that causality implies a polynomial bound in the upper-half $s$ plane, for $t<0$, given certain assumptions.  This is naturally associated with 
scattering producing a time {\it delay} (causal) rather than a time {\it advance} (acausal), as is most simply seen by considering $0+1$ - dimensional scattering, with initial and final amplitudes related by 
\beq
\psi_f(t)=\int_{-\infty}^\infty dt'S(t-t') \psi_i(t')\ .
\eeq
By causality, if the source $\psi_i$ vanishes for $t'<0$, the response $\psi_f$ should as well.  This is achieved if  the Fourier transform $S(E)$ is polynomially bounded in the upper half $E$ plane.

\ex{Check this last statement}

\begin{figure}[!hbtp]
\begin{center}
\includegraphics[width=8cm]{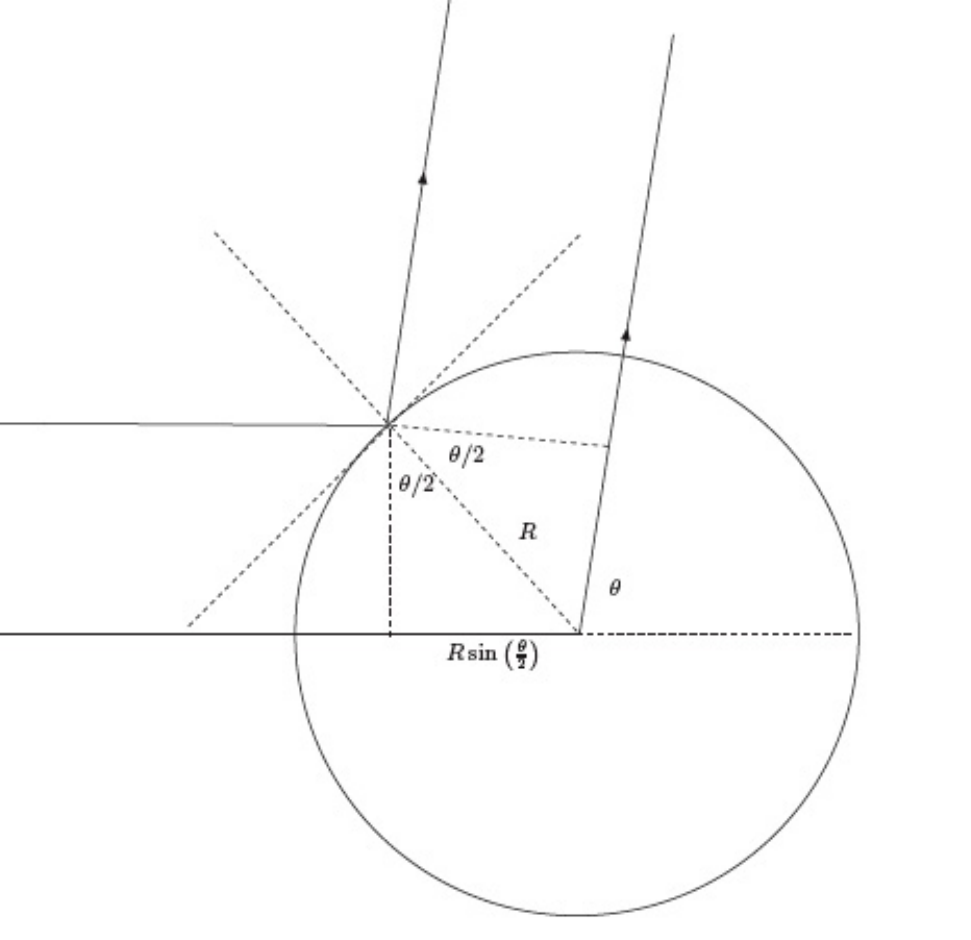}
\caption{Scattering from a repulsive potential with range $R$ produces a time advance proportional to $R$.}\label{fig:repuls}
\end{center}
\end{figure} 

In higher-dimensional scattering, there is a simple argument that repulsive scattering with range growing as a power of the energy produces nonpolynomially-bounded behavior, associated with a time advance. Specifically, consider fig.~\ref{fig:repuls}.  For scattering at angle $\theta$, the path that scatters off the potential at $R(E)\propto E^p$ is shortened compared to that through the origin, corresponding to no scattering.  Simple geometry gives the time advance, and its corresponding phase shift, yielding
\beq
S\sim e^{- i \sqrt{-t} R(E)}\ .
\eeq
which violates polynomial boundedness in the upper-half $s$ plane.  

\begin{figure}[!hbtp]
\begin{center}
\includegraphics[width=8cm]{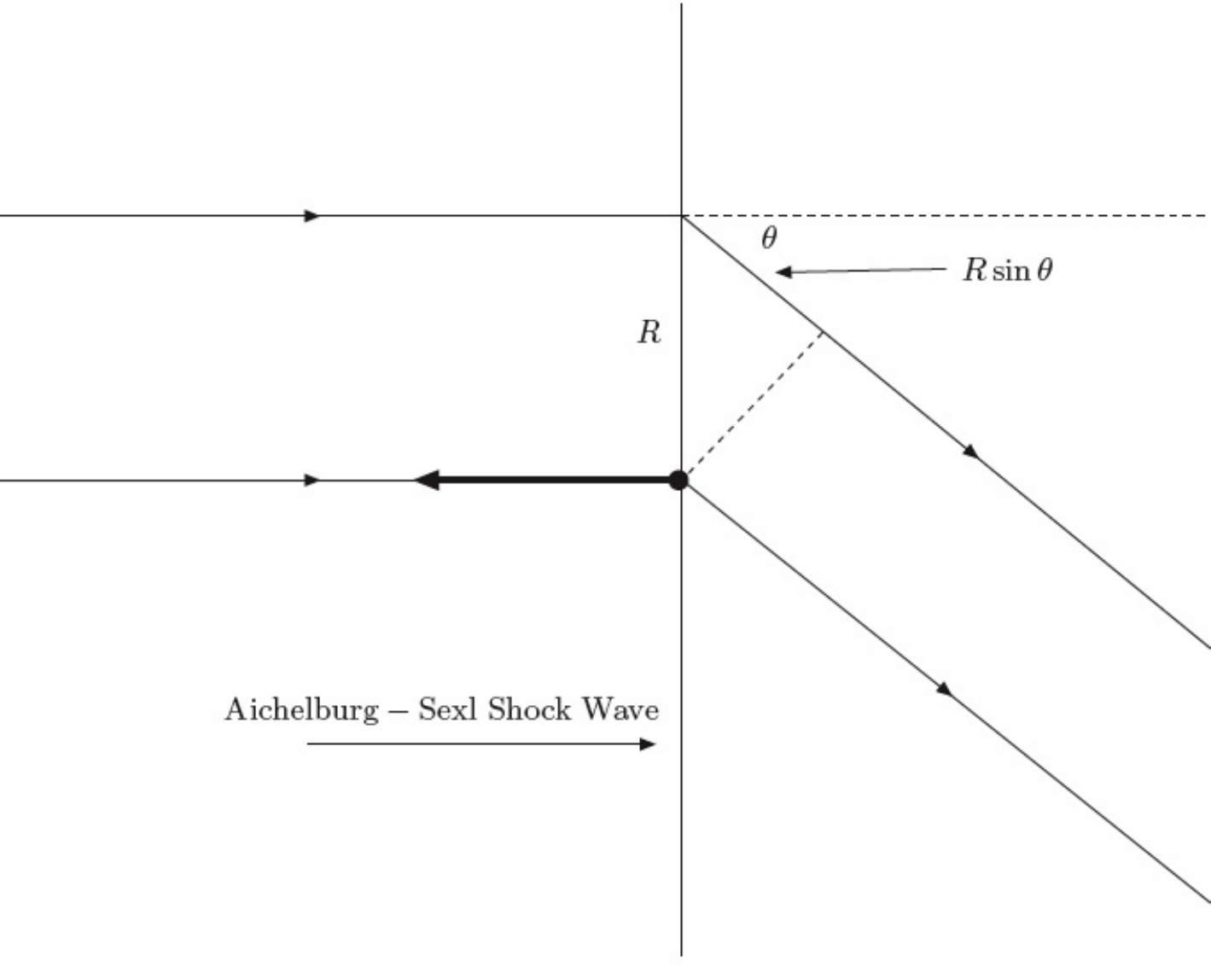}
\caption{A schematic picture of (attractive) scattering from an Aichelburg-Sexl shock, producing a lengthened path, therefore a time delay.}\label{fig:ASscatt}
\end{center}
\end{figure} 

While I have stressed the long-range nature of gravity, gravity is also {\it attractive}, indicating a different story.  For example, consider instead a picture of scattering of a particle from an Aichelberg-Sexl shock, shown in fig.~\ref{fig:ASscatt}.  The attractive behavior produces a time delay, and we now have polynomially-bounded behavior,
\beq\label{causscatt}
S\sim e^{ i \sqrt{-t} R(E)}\ .
\eeq

Nonetheless, behavior such as \eqref{causscatt} is nonpolynomial -- its non-polynomial growth is just in a different region than Im$(s)>0$, $t<0$. 
This causal yet nonpolynomial behavior is the kind suggested by our arguments in the preceding subsection.
 So, if we think of locality as characterized in terms of nonpolynomiality of the S-matrix, these arguments appear to hang together, and  
  suggest a sense in which scattering in gravity could walk a fine  line of {\it nonlocal} and {\it causal} behavior, thus offering a possible outline for a consistent story.  I have really only given an outline of how such arguments could be made.  Further discussion of how such ideas could be part of the story of black hole evaporation is given in \cite{GiNLvC}.

Finally, there is a slightly different way of describing locality/causality in the S-matrix framework; see {\it e.g.} Coleman's 1969 lectures\cite{Coleman:1969xz} in this school.  Specifically, one considers two subsets of scattered particles, in the limit as the separation between the subsets is taken large.  {\it Cluster decomposition} is the statement that the leading term in the limit of large separation is that where the two sets of particles scatter independently.  {\it Macrocausality} is the statement that the next most important terms arise from a single particle from one of the subsets scattering with the other subset. If we consider gravity, we find that such statements are expected to be true, but in a more limited region of validity than in other field theories.  

Specifically, this arises from an order of limits question.  If we fix a large separation $|x-x'|$ between the two clusters of particles, we expect cluster decomposition and macrocausality to break down when the CM energy of the two clusters,  $|k+k'|$, becomes large enough to violate the locality bound \eqref{Locbd}.  In fact, as we saw in section \ref{Pertquant}, we expect other important contributions even for 
\beq
|x-x'|^{D-4}\sim G_D |k+k'|^2\ .
\eeq
There is still however a more limited sense in which clustering and macrocausality hold; for fixed CM energy, both of these statements are expected to be true in gravity for sufficiently large separation $|x-x'|$.  So, part of the question is how to achieve this degree of locality, yet modify microcausality \eqref{Opcomm}, and plausibly the very existence of a description in terms of quantum field states, in contexts where the locality bound is violated.

\section{Conclusions and directions}

There are many facets of the story of quantum gravity; some future directions are indicated by incompleteness, unproven statements, and exercises in these notes.  More broadly, there are several  important and general themes for investigation.  

While there has historically been a lot of focus on the nonrenormalizability and singularities of gravity, I have argued that {\it unitarity} is a more profound problem in formulating a complete theory.  This is particularly revealed in gedanken experiments involving ultrahigh-energy scattering. These notes have given sharpened statements that this unitarity crisis is a {\it long-distance} issue, and there is no clear path to its resolution in short-distance alterations of the theory.  This appears to be  a critically important and central aspect in the problem of quantizing gravity, suggesting that these considerations are an important guide towards the shape of the ultimate theory.

In investigating these questions, one seeks sharp physical quantities that a theory of gravity should be able to compute.  One possibility is quantum mechanical observables that approximate local observables in certain states and circumstances.  Another, in cases where, as we appear to observe, there is a good flat space limit, is the gravitational S-matrix.

In particular, investigation of ultrahigh-energy scattering, specifically in the black-hole regime, reveals deep conflict between the ``local" approach to describing physics and a more ``asymptotic" or scattering approach.  This appears to indicate a failure or modification of the ``local" approach even at macroscopic distances, under the appropriate circumstances, which does not appear to be curable by short-distance modifications of the theory.  A profound question is how to reconcile such failure with the use and validity of local quantum field theory under ordinary circumstances.  

While specific frameworks for quantum gravity have been proposed, they do not yet satisfactorily resolve these problems.  Loop quantum gravity is still grappling with the problem of approximating flat space and producing an S-matrix.  Despite initial promise, string theory has not yet advanced to the stage where it directly addresses the tension between the asymptotic and local approaches, or is able to compute a unitary S-matrix in the relevant strong gravity regime.  Because of the long-distance and non-perturbative nature of the problem, it is also not clear how it would be addressed if other problems of quantum gravity were resolved, for example if supergravity indeed yields perturbatively finite amplitudes\cite{BDR}.

I have argued that the needed theory should have certain non-local features, yet be nearly-local, to avoid consistency problems associated with acausality, and also since we expect clustering and macrocausality to hold in certain limits.  

One way to seek further information is to explore the interplay between the local and asymptotic descriptions, and the limitations of the former.  Cosmology, particularly inflationary cosmology, provides another testing ground for this, whose discussion would require another set of lectures.  In particular, one can explore the formulation and role of relational and protolocal observables\cite{GMH,SGMadS}, and apparent limitations to a perturbative approach\cite{QBHB,GiSlone,GiSltwo}.   One can look for other clues by investigating the nature of the nonperturbative dynamics, and the structure of a quantum framework\cite{UQM}, needed to give a more complete treatment of cosmology or of other dynamical spacetimes such as black holes. 

Further clues can also be sought directly through the S-matrix.  We know that such a matrix capable of describing gravity must have some rather special properties, in addition to general properties such as unitarity and analyticity.  Particularly, we have seen that crossing becomes more subtle, and that there are interesting questions surrounding asymptotic behavior.  Unitarization of the S-matrix in the strong gravity regime is a key question.  It has been suggested that one can have such an S-matrix consistent with causality, but with modifications of the usual indicators of locality.  

The bigger question is what mechanisms, principles, and framework underly the theory.  Particularly, we seem to need to relax or modify locality such that it is not a fundamental property, but emerges approximately in appropriate circumstances, and moreover to avoid inconsistencies usually associated with acausality -- this seems a critical tension.  This ``locality without locality" is plausibly a powerful constraint on the ultimate shape of such a ``nearly-local" mechanics.

\vskip.2in
\noindent{\bf Acknowledgements} I would like to thank the organizers for the invitation to present these lectures at Erice.  I have greatly benefited from discussions with my collaborators J. Andersen, M. Gary, D. Gross, J. Hartle, M. Lippert, A. Maharana, D. Marolf, J. Pendeones, R. Porto, M. Schmidt-Sommerfeld, M. Sloth, M. Srednicki, as well as with N. Arkani-Hamed, B. Grinstein, and E. Witten; however, none of them should be blamed for misconceptions, which are my own. 
This work is supported in
part by the U.S. Dept. of Energy under Contract
DE-FG02-91ER40618, and by grant FQXi-RFP3-1008 from the Foundational Questions Institute (FQXi)/Silicon Valley Community Foundation.

\end{document}